\documentclass[aps,twocolumn,prb,preprintnumbers,amsmath,amssymb,superscriptaddress,floats]{revtex4}

\usepackage{graphicx}
\usepackage{bm}
\usepackage{hyperref}
\usepackage{natbib}
\usepackage{braket}
\usepackage{float}
\usepackage{color}
\usepackage{nicefrac}

\begin{document}

\title{Low energy magnon dynamics and magneto-optics of the skyrmionic Mott insulator Cu$_2$OSeO$_3$}

\author{N. J. Laurita}
\affiliation{Institute for Quantum Matter and Department of Physics and Astronomy, The Johns Hopkins University, Baltimore, MD 21218, USA}

\author{G. G. Marcus}
\affiliation{Institute for Quantum Matter and Department of Physics and Astronomy, The Johns Hopkins University, Baltimore, MD 21218, USA}

\author{B. A. Trump}
\affiliation{Institute for Quantum Matter and Department of Physics and Astronomy, The Johns Hopkins University, Baltimore, MD 21218, USA}
\affiliation{Department of Chemistry, Johns Hopkins University, Baltimore, Maryland, 21218, USA}

\author{J. Kindervater}
\affiliation{Institute for Quantum Matter and Department of Physics and Astronomy, The Johns Hopkins University, Baltimore, MD 21218, USA}

\author{M. B. Stone}
\affiliation{Quantum Condensed Matter Division, Oak Ridge National Laboratory, Oak Ridge, Tennessee 37831, USA}

\author{T. M. McQueen}
\affiliation{Institute for Quantum Matter and Department of Physics and Astronomy, The Johns Hopkins University, Baltimore, MD 21218, USA}
\affiliation{Department of Chemistry, Johns Hopkins University, Baltimore, Maryland, 21218, USA}
\affiliation{Department of Materials Science and Engineering, Johns Hopkins University, Baltimore, Maryland, 21218, USA}

\author{C. L. Broholm}
\affiliation{Institute for Quantum Matter and Department of Physics and Astronomy, The Johns Hopkins University, Baltimore, MD 21218, USA}
\affiliation{Department of Materials Science and Engineering, Johns Hopkins University, Baltimore, Maryland, 21218, USA}

\author{N. P. Armitage}
\affiliation{Institute for Quantum Matter and Department of Physics and Astronomy, The Johns Hopkins University, Baltimore, MD 21218, USA}

\date{\today}

\begin{abstract}
\noindent
In this work, we present a comprehensive study of the low energy optical magnetic response of the skyrmionic Mott insulator Cu$_2$OSeO$_3$ via high resolution time-domain THz spectroscopy.  In zero field, a new magnetic excitation not predicted by spin-wave theory with frequency $f$ =  2.03 THz is observed and shown, with accompanying time-of-flight neutron scattering experiments, to be a zone folded magnon from the $\mathrm{R}$ to $\mathrm{\Gamma}$ points of the Brillouin zone.  Highly sensitive polarimetry experiments performed in weak magnetic fields, $\mu_0$H $<$ 200 mT, observe Faraday and Kerr rotations which are proportional to the sample magnetization, allowing for optical detection of the skyrmion phase and construction of a magnetic phase diagram. From these measurements, we extract a critical exponent of $\beta$ = 0.35 $\pm$ 0.04, in good agreement with the expected value for the 3D Heisenberg universality class of $\beta$ = 0.367.  In large magnetic fields, $\mu_0$H $>$ 5 T, we observe the magnetically active uniform mode of the ferrimagnetic field polarized phase whose dynamics as a function of field and temperature are studied.  In addition to extracting a $g_\text{eff}$ = 2.08 $\pm$ 0.03, we observe the uniform mode to decay through a non-Gilbert damping mechanism and to possesses a finite spontaneous decay rate, $\Gamma_0$ $\approx$ 25 GHz, in the zero temperature limit.  Our observations are attributed to Dzyaloshinkii-Moriya interactions, which have been proposed to be exceptionally strong in Cu$_2$OSeO$_3$ and are expected to impact the low energy magnetic response of such chiral magnets.
\end{abstract}

\maketitle

\section{Introduction}

Non-trivial spin textures have become a hotbed of research due to their unique physical properties and potential applications in spintronics and information storage.  Skyrmions \cite{Skyrme1962}, topological whirls of magnetic spins, are a prime example of such a non-trivial spin texture \cite{Nagaosa2013}.  A skyrmion phase, in which a hexagonal lattice of skyrmions is formed, was recently predicted to exist in chiral magnets \cite{Bogdanov1989, Rozler2006}, and has since been observed in the metallic B20 helimagnets: MnSi \cite{Muhlbauer2009}, FeGe \cite{Yu2011}, and Fe$_{1-x}$Co$_x$Si \cite{Yu2010, Munzer2010}.  The skyrmion phases of these materials possess unique electrodynamics \cite{Schultz2012} such that they can be manipulated via application of an electrical current or thermal gradient \cite{Jonietz2010} and have accordingly attracted intense experimental interest.

Optical spectroscopy is exceptionally well-suited for studying these chiral magnets as their nearly ferromagnetic nature ensures an ordering wavevector of $\vec{k}$ $\approx$ 0 which is directly probed by optical experiments.  However, high precision optical transmission experiments of single crystal skyrmion materials has thus far been impossible due to their metallic nature. This has been particularly detrimental to the optical study of their low energy magnetic response which generally requires bulk samples. 

Recently, a skyrmion phase with unique physical properties was shown to exist in the \textit{insulating} chiral magnet Cu$_2$OSeO$_3$ \cite{SekiA2012, SekiB2012, White2012}.  The low symmetry crystal structure of Cu$_2$OSeO$_3$ permits multiferroism \cite{Yang2012, Khomskii2009, Tokura2014} as well as magnetoelectric coupling \cite{Jia2006, Jia2007, Arima2007}, which recent measurements have shown results in a finite polarization that onsets in conjunction with magnetic order at T$_c$ $\approx$ 58K \cite{Bos2008, Belesi2012}.  This finite polarization allows for coupling between \textit{magnetic} skyrmions and applied \textit{electric} fields \cite{SekiB2012, White2012} - a promising mechanism for technological applications and novel devices \cite{Mochizuki2013, Mochizuki2012}.  Accordingly, the magnetic and magnetoelectric properties of Cu$_2$OSeO$_3$ have been the focus of intense investigation \cite{Bos2008, Maisurandze2012, Maisurandze2011, Omrani2014, Belesi2012, Ruff2015}.  

From an optics perspective, the large Mott insulating gap of Cu$_2$OSeO$_3$ naturally separates electric and magnetic degrees of freedom \cite{Janson2014, Miller2010, Versteeg2016}, allowing for direct access to the magnetic response via transmission optics. Spectroscopic investigations have since been performed from the microwave \cite{Szaller2013, Okamura2015, Mochizuki2015, Schwarze2015} to the visible \cite{Versteeg2016} frequency ranges.  However, experiments performed at infrared or terahertz (THz) frequencies have so far only occurred at two extremes of the phase diagram, either in zero applied magnetic field \cite{Miller2010, Gnezdilov2010} or in large pulsed magnetic fields of order $\mu_0$H $\approx$ 10T \cite{Ozerov2014}.  To date, no THz experiments have been performed in weak magnetic fields ($\mu_0$H $\leq$ 200 mT), within the various magnetic phases of Cu$_2$OSeO$_3$, including the skyrmion phase.  Additionally, a detailed investigation into the dynamics of the known THz excitations as a function of temperature and magnetic field has not yet been presented.  

In this work, we present a comprehensive high resolution optical study of the skyrmion insulator Cu$_2$OSeO$_3$ in the THz regime via time-domain THz spectroscopy (TDTS).  As our experimental energy range, $\hbar \omega$ = 1 - 10 meV, is far less than the bulk band gap, $\Delta _g \approx$ 2 eV \cite{Versteeg2016}, we directly access the low energy magnetic response of Cu$_2$OSeO$_3$.  Experiments are performed within three distinct regimes of magnetic field: $\mu_0$H = 0, $\mu_0$H $\leq$ 200 mT, and $\mu_0$H $\ge$ 5 T.  In zero field, we observe a new magnetic excitation which is revealed to be a zone folded magnon from the zone boundary to the zone center which has not been predicted by spin-wave theory.  Highly sensitive polarimetry experiments performed in weak magnetic fields observe Faraday and Kerr rotations which are proportional to the sample magnetization, allowing for optical detection of the skyrmion phase and construction of a magnetic phase diagram.  In large magnetic fields, we study the field and temperature dependent dynamics of the uniform mode of the field polarized phase.  The uniform mode is found to decay through a non-Gilbert damping mechanism and to possess a finite spontaneous decay rate in the zero temperature limit.  The potential damping mechanisms of this mode are discussed.  

\section{H-T Phase Diagram}

\begin{figure}
\includegraphics[width=1.0\columnwidth, keepaspectratio]{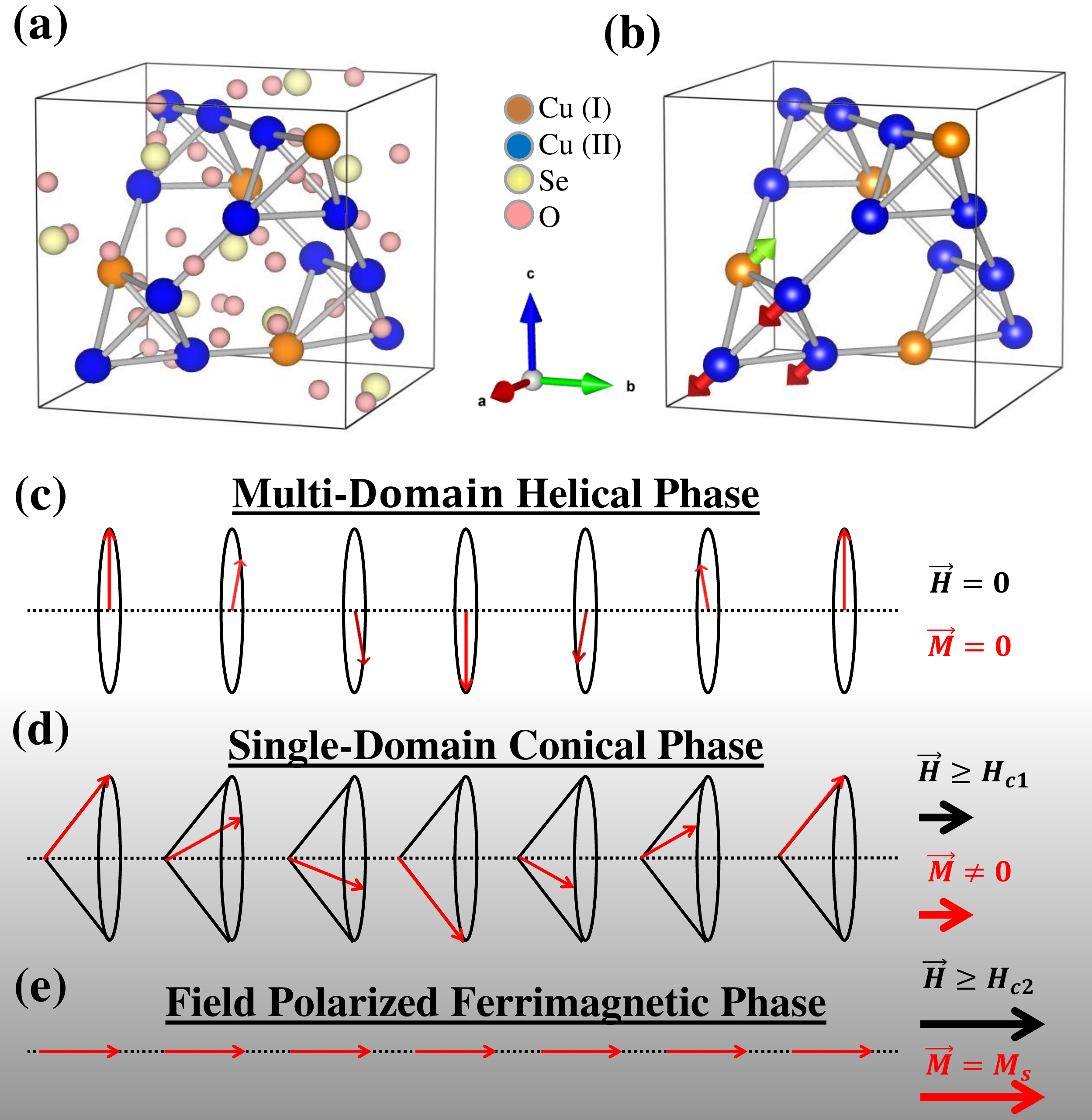}
\caption{(a) Unit cell of Cu$_2$OSeO$_3$ with the Cu(I) and Cu(II) positions shown as orange and blue spheres respectively.  (b) The ground state of each tetrahedron consists of a ferrimagnetic arrangement in which the Cu(I) spin (green arrow) orders antiferromagnetically to the Cu(II) spins (red arrows), creating effective S=1 spins.  Shown in (c)-(e) are representations of the (b) helical, (c) conical, and (d) field polarized magnetic phases where each arrow represents the effective spin of a single tetrahedron.  See text for more details.}
\label{Fig1}
\end{figure}

\autoref{Fig1}(a) shows the crystal structure of Cu$_2$OSeO$_3$ \cite{Momma2011} which crystallizes in the cubic, but non-centrosymmetric, space group P2$_1$3 \cite{Effenberger1986}.  The unit cell forms a distorted pyrochlore lattice with 16 Cu$^{2+}$ (S = 1/2) ions residing on the vertices of 4 corner sharing tetrahedra.  Each tetrahedron is composed of one Cu(I) site (orange spheres) and three Cu(II) sites (blue spheres), which possess distinct crystal field environments \cite{Bos2008, Belesi2010}.  This low symmetry structure results in five unique Heisenberg exchange interactions and five Dzyaloshinskii-Moriya (DM) \cite{Dz1958, Moriya1960} interactions within the unit cell.  These exchange interactions are classified as either ``strong" or ``weak'' depending on whether they couple two intra-tetrahedral or two inter-tetrahedral Cu$^{2+}$ spins respectively.  Experiments \cite{Portnichenko2016, Tucker2016} and calculations \cite{Janson2014, Romhanyi2014, Chizikov2015} reveal that the ``strong" couplings result in a semi-classical ferrimagnetic arrangement for each tetrahedron, in which the Cu(I) spin orders antiferromagnetically to the three parallel Cu(II) spins (\autoref{Fig1}(b)). This ground state is well separated from the first excited state by a large energy gap of $\Delta$ $\approx$ 275K, \cite{Portnichenko2016, Tucker2016, Janson2014, Romhanyi2014} such that each tetrahedron can be treated as an effective S = 1 spin - which form the basic magnetic building blocks of Cu$_2$OSeO$_3$.   The resulting effective unit cell then consists of four S = 1 spins arranged in the Trillium lattice, a structure identical to that of the B20 helimagnets, revealing why such similar phase diagrams result from seemingly dissimilar compounds \cite{Schwarze2015}.

The magnetic phase diagram can then be understood as competition between the ``weak'' Heisenberg  and ``weak'' DM exchange interactions between these effective S = 1 spins \cite{Janson2014, Romhanyi2014, Chizikov2015}.  With $|{D_{ij}}|$ $< |J_{ij}|$, the resultant magnetic order in zero field onsets at T$_c$ $\approx$ 58K in the form of a long wavelength helix ($\lambda$ $\approx$ 50 nm) \cite{SekiA2012, Adams2012} as shown in \autoref{Fig1} (c).  The magnetic order reduces the symmetry to the rhombohedral group R3 \cite{Bos2008}.  Weak cubic anisotropy pins these helices to degenerate high symmetry directions of the cubic structure, resulting in a``multi-domain helical phase."  Application of a magnetic field cants the spins in the direction of the applied field.  At H$_{c1}$ the applied field overcomes the weak cubic anisotropy resulting in a ``single-domain conical phase," shown in \autoref{Fig1}(d), in which the helices co-align into a single domain with a conical arrangement of spins \cite{SekiA2012, Adams2012}.  Further increasing the applied magnetic field smoothly tunes the cone angle to zero at H$_{c2}$, thereby untwisting the magnetization, resulting in a field polarized ferrimagnetic phase as shown in \autoref{Fig1}(e).  While the exact values of the critical fields depends on the demagnetization factors of the sample, H$_{c1}$ and H$_{c2}$ are generally on the order of 10 mT and 100 mT respectively \cite{SekiA2012, Adams2012}.

Much like the B20 helimagnets, a skyrmion phase spanned by $\approx$ 2K and $\approx$ 30 mT just below T$_c$ is stabilized by Gaussian thermal fluctuations \cite{Muhlbauer2009} and has since been detected by a variety of techniques \cite{Adams2012, White2012, SekiB2012, SekiA2012, Levatic2014, Omrani2014}.  In this phase, skyrmions form a hexagonal lattice much like Abrikosov vortices in type-II superconductors.  Such a phase can be thought of as a double twisting of the magnetization which results from the superposition of 3 helices with $\vec{k}$ vectors at 120 degrees to one another \cite{Adams2012}.  The skyrmion diameter is identical to the helical phase wavelength, d $\approx$ 50 nm, which is three orders of magnitude larger than the inter-atomic spacing \cite{SekiA2012} revealing skyrmions to be vast mesoscopic spin structures.

\section{Methods}

Phase pure single crystals of Cu$_2$OSeO$_3$ were grown by chemical vapor transport. Cu$_2$OSeO$_3$ powder was placed in an evacuated fused-silica tube with a temperature gradient of 640 $^{\circ}$C - 530 $^{\circ}$C, with NH$_4$Cl as the transport additive, using seed crystals to increase yield. Purity of single crystals was verified by magnetization and X-ray diffraction experiments, showing reproducibility of physical property behavior and good crystallinity. For more details see Ref. \onlinecite{Panella2017}.  

TDTS measurements were performed on a hand polished single crystal sample with plane parallel sides of cross sectional area of $\approx$ 3 mm $\times$ 3 mm and thickness $d$ = 0.92 mm.  The orientation of the sample was such that the (1,$\bar{1}$,0) direction was normal to the sample surface.  Experiments were performed using a home built spectrometer with applied magnetic fields up to 7 T in Faraday geometry ($\vec{\text{k}}_{\text{THz}}$ $\parallel$ $\vec{\text{H}}_{\text{dc}}$) \cite{Laurita2016}.  The polarization was such that the THz oscillatory fields $\bf{e_\text{ac}}$ $\parallel$ $\bf{c}$ and $\bf{h_\text{ac}}$ $\parallel$ (110) directions respectively.  TDTS is a high resolution method for accurately measuring the electromagnetic response of a sample in the experimentally challenging THz range.  In a typical TDTS experiment, the electric field of a THz pulse transmitted through a sample is measured as a function of real time. Fourier transforming the measured electric field and referencing to an aperture of identical size allows access to the frequency dependent \textit{complex} transmission spectrum of the sample which is given by,
\begin{equation}
\widetilde{T} = \frac{4\widetilde{n}}{(\widetilde{n}+1)^2}\exp{(\frac{i \omega d}{c}(\widetilde{n}-1))}.
\label{Transeq}
\end{equation}

\noindent Here $d$ is the sample thickness, $\omega$ is the frequency, $c$ is the speed of light, $\widetilde{n}$ is the sample's complex index of refraction, and normal incidence has been assumed. A Newton-Raphson \cite{Press2007} based numerical inversion of the complex transmission is then used to obtain both the frequency dependent real and imaginary parts of the index of refraction.  

The index of refraction, $\widetilde{n} = \sqrt{\epsilon \mu}$ = $n + ik$, contains both the electric and magnetic responses of the sample as THz fields can couple to both electric and magnetic dipole transitions.  In principle, the linear magnetoelectric properties of Cu$_2$OSeO$_3$ introduce an additional contribution to the index of refraction such that $\widetilde{n} = \sqrt{\epsilon \mu \pm \chi ^{\text{ME}}}$, where $\chi ^{\text{ME}}$ is the magnetoelectric susceptibility.  However, at the level of sensitivity of the present experiments we observe no magnetoelectric effects in the THz range suggesting that the magnetoelectric susceptibility is small compared to the linear electric and magnetic susceptibilities, $\chi ^{\text{ME}}$ $\ll$ $\chi ^{\text{M}},\chi ^{\text{E}}$.  We therefore neglect the magnetoelectric contribution to the index of refraction in our analysis and ascribe absorptions as stemming from purely electric or magnetic effects.

The linear THz response of a sample can be represented in the Jones calculus \cite{Jones1941} as a 2$\times$2 complex transmission matrix of the form,
$$\hat{T} = \begin{bmatrix} T_{xx}&T_{xy} \\ T_{yx}&T_{yy} \end{bmatrix} $$
However, the overall symmetry of Cu$_2$OSeO$_3$ restricts the response such that the transmission matrix is fully antisymmetric, i.e. $T_{xx} = T_{yy}$ and $T_{xy} = -T_{yx}$ \cite{Armitage2014}. One can then identify off diagonal elements of the transmission matrix with rotation of the plane of polarization of light by the sample.  Polarization rotation experiments were done through the use of a rotating polarizer \cite{Morris2012} technique, which allows for simultaneous measurement of two elements of the transmission matrix. The complex rotation angle is then given by the relation $\theta = \tan^{-1}({\frac{T_{yx}}{T_{xx}}})$.  Fully antisymmetric transmission matricies can be diagonalized by a circular basis transformation, $T_r$ = $T_{xx}$ - $iT_{xy}$ and $T_l$ = $T_{xx}$ + $iT_{xy}$, suggesting experiments performed in Faraday geometry are best understood in the circular basis.  Data taken in applied magnetic field will therefore be presented as either a polarization rotation or in the circular basis. 

Time-of-flight neutron scattering experiments were performed on the SEQUOIA instrument at the Spallation Neutron Source of Oak Ridge National Laboratory. To enhance the signal to noise ratio, we co-aligned more than 50 single crystals to yield a mass of $\approx$ 5g. A mosaic of less than 0.5 degrees was ensured by design of a custom mount to orient the samples according to their as-grown facets.  The co-aligned mosaic was cooled to 4~K in a bottom-loading CCR.  An incident energy of 20~meV was chosen with the fine chopper rotating at a rate of 180~Hz.  Individual monochromatic measurements were performed as the sample was rotated through 180 degrees in 0.5 degree steps about the ($h$ $\bar{h}$ 0) axis.  These same spectrometer settings were used to measure Vanadium incoherent scattering for absolute normalization of the differential scattering cross-section. Reduction of the data was performed using Mantid \cite{Arnold2014} and subsequent analysis was performed with Horace \cite{Ewings2016}.

\section{Experimental Results}
\subsection{Temperature Dependence}

\begin{figure}
\includegraphics[width=1.0 \columnwidth, keepaspectratio]{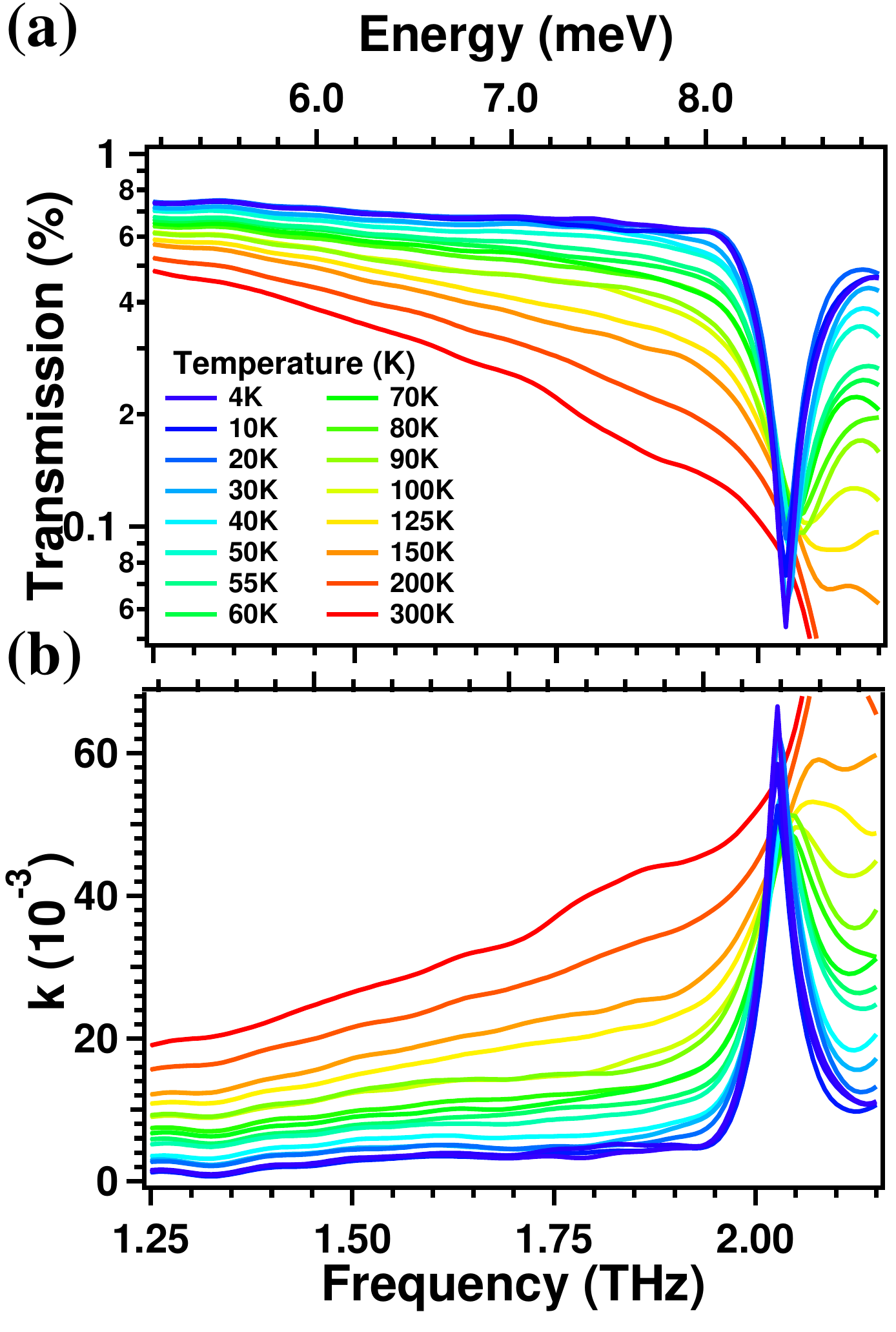}
\caption{(a) Magnitude of the zero field complex transmission spectra of Cu$_2$OSeO$_3$ as a function of frequency and temperature plotted on log scale.  (b) Corresponding imaginary, or dissipative, part of the index of refraction extracted from the transmission and Eq. \ref{Transeq}.  A clear absorption is observed at $f_0$ $\approx$ 2.03 THz which narrows and gains intensity with reducing temperature.}
\label{Fig2}
\end{figure}

\autoref{Fig2}(a) displays the magnitude of the complex zero field transmission of Cu$_2$OSeO$_3$ as a function of frequency and temperature plotted on log scale.  \autoref{Fig2}(b) displays the corresponding imaginary, or dissipative, part of the index of refraction extracted from the transmission and Eq. \ref{Transeq}.
One can see that the spectra consists of two prominent features,   The first is a nearly linear background which shows decreasing dissipation as the temperature is reduced.  The origin of this background is an intense infrared active phonon at 2.5 THz, which is outside our experimental frequency range \cite{Miller2010}.  The reduction of this background with reducing temperature presumably results from a narrowing of this phonon at lower temperatures.  The second, and more interesting, feature in the spectra is the clear absorption with a resonant frequency of $f_0$ = 2.03 THz (8.40 meV).  As shown in \autoref{Fig2}, this excitation begins developing at T $\approx$ 120 K and displays an increasing intensity and simultaneous narrowing as the temperature is reduced. 

This absorption was previously reported in far infrared experiments performed by Miller \textit{et al.} \cite{Miller2010} in which it was hypothesized to be a low frequency phonon.  It was reported that this absorption displayed no response to weak magnetic fields up to 14 mT applied parallel to the sample surface and no anomalous behavior at the magnetic ordering temperature T$_c$ $\approx$ 58K.  Although the potential magnetic dipole or magnetoelectric character of the excitation could not be excluded as the response to larger magnetic fields or anisotropy upon change in field direction or incident polarization was not investigated.  

However, the intensity of this excitation is generally problematic for the phonon interpretation.  This becomes obvious when one compares the spectral weight (plasma frequency) of this excitation to that of the other known infrared optical phonons of Cu$_2$OSeO$_3$, which were also reported by Miller \textit{et al.} \cite{Miller2010}.  In general, a phonon's plasma frequency can be related to its total spectral weight through the sum rule $\int_0^\infty \sigma _1^{\text{phonon}}(\omega)\propto \omega _p^2$.  A comparison reveals that the spectral weight of the low frequency excitation observed in this work is a staggering 10$^4$ to 10$^8$ times weaker than any of the other infrared optical phonons observed in Cu$_2$OSeO$_3$, suggesting a different origin for this excitation.  Instead, the intensity of this excitation is much more consistent with magnetic excitations in single crystal samples.  The weak intensity of magnetic excitations derives from the fact that the THz magnetic field interacts far more weakly with matter than the THz electric field.

\begin{figure}
\includegraphics[width=0.85\columnwidth, keepaspectratio]{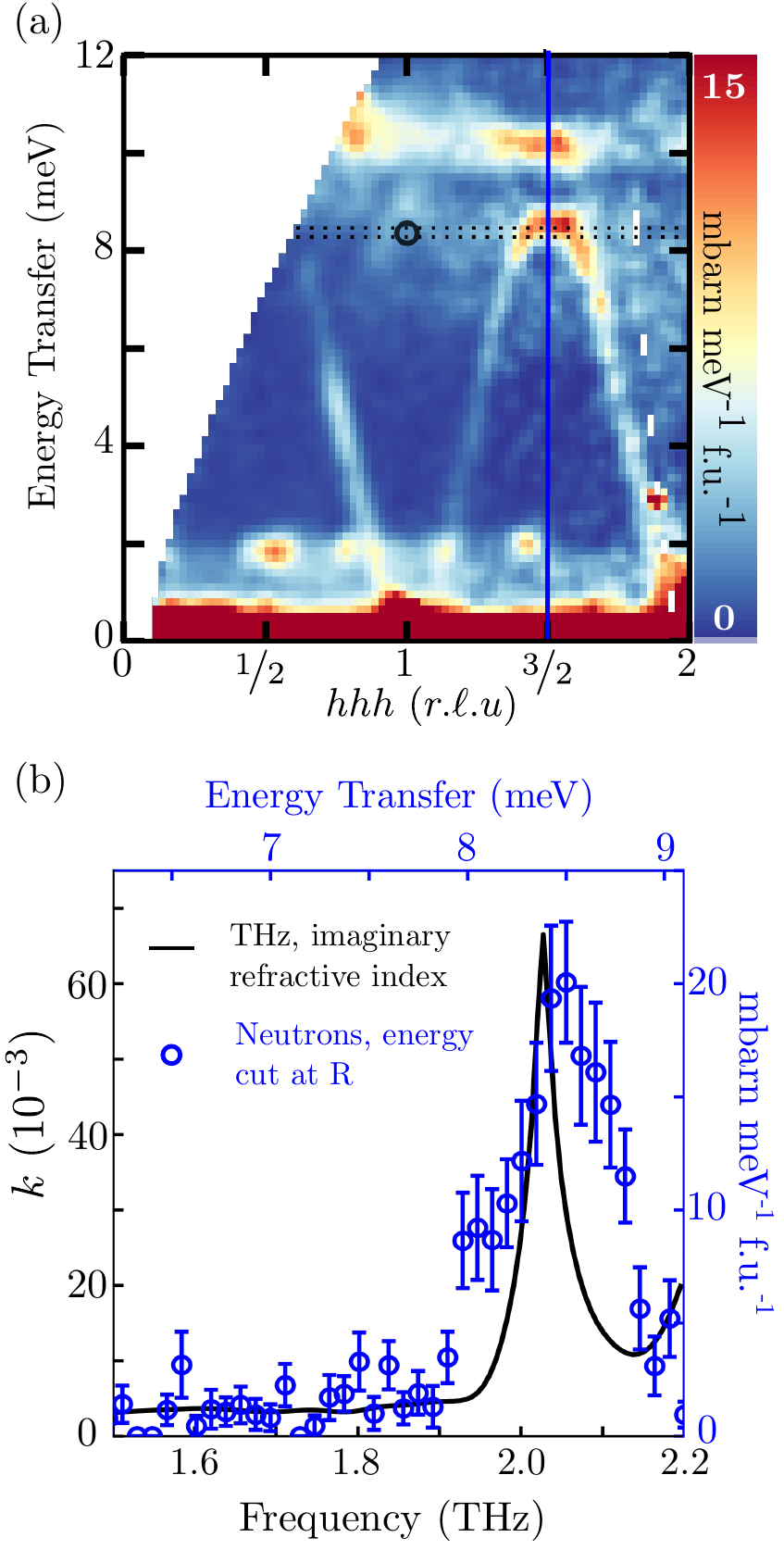}
\caption{(a) $\mathbf{Q}$-$E$ dependence of the differential scattering cross-section along the (111) direction in reciprocal space. Perpendicular directions have been integrated within 0.1~\AA$^{-1}$. The magnon band at $\mathrm{R}_{111}$ is seen to be folded to the $\Gamma_{111}$ zone-center. The peak (black circle) and FWHM (black dotted lines) in the THz spectrum are overlaid for comparison. 
(b) An energy cut (blue, right axis) along the blue line shown in (a) overlaid on the imaginary part of the index of refraction (black, left axis) measured by THz spectroscopy.  All data is obtained at T= 4.0(5)~K.}
\label{THzNeutron}
\end{figure}

Further support for the magnon interpretation of this excitation is provided by the momentum resolved capabilities of inelastic neutron scattering.  Shown in \autoref{THzNeutron}(a) is a false color map of the differential scattering cross-section at 4~K (to be detailed in a related upcoming publication \cite{Marcus2017}).  The peak energy of the excitation in question has been overlaid at the (111) zone-center ($\Gamma_{111}$), where a  dispersive magnon branch reaches its highest point.  Judging from its energy, relatively weak intensity, and local curvature at the apogee, it appears to be a zone-folded replica of the magnon band whose intensity is strongly peaked at the zone boundary  $\mathrm{R}_{111}=(\nicefrac{3}{2}\, \nicefrac{3}{2}\, \nicefrac{3}{2})$ point.  This observation suggests a more direct comparison between these two spectroscopic techniques.

\autoref{THzNeutron}(b) shows an energy cut at R$_{111}$, along the blue line in \autoref{THzNeutron}(a), overlaid with the dissipative part of the index of refraction from THz spectroscopy.  This presentation makes evident the agreement in energy between these modes, which is determined by neutron scattering as $f$=2.05(3)~THz.  Differences in the excitation width result from the inherent energy resolution limit of the neutron spectrometer.  The calculated full width at half maximum (FWHM) instrumental energy resolution of the neutron spectrometer is 0.31 meV at the energy transfer of the observed excitation.  Together, these observations motivate the conclusion that the mode observed by THz spectroscopy is, in fact, a zone-folded magnon---a new magnetic excitation which has not been previously predicted by spin-wave theory \cite{Janson2014, Romhanyi2014}.   

With the magnetic character of this excitation determined, the dynamical properties of this magnon can be found from fitting the spectra to an oscillator model with the following form,
\begin{equation}
\mu(\omega) = \frac{S \omega _0 ^{2}}{\omega _0^2 - \omega ^{2} - i \omega \Gamma } + \mu _{\infty}  
\label{Eq1}
\end{equation}
where, $\omega _0$, $\Gamma$, $S$, and $\mu_\infty$ represent the magnon frequency, full width at half max, oscillator strength, and high frequency permeability of the lattice respectively.  

\begin{figure}
\includegraphics[width=1.0\columnwidth, keepaspectratio]{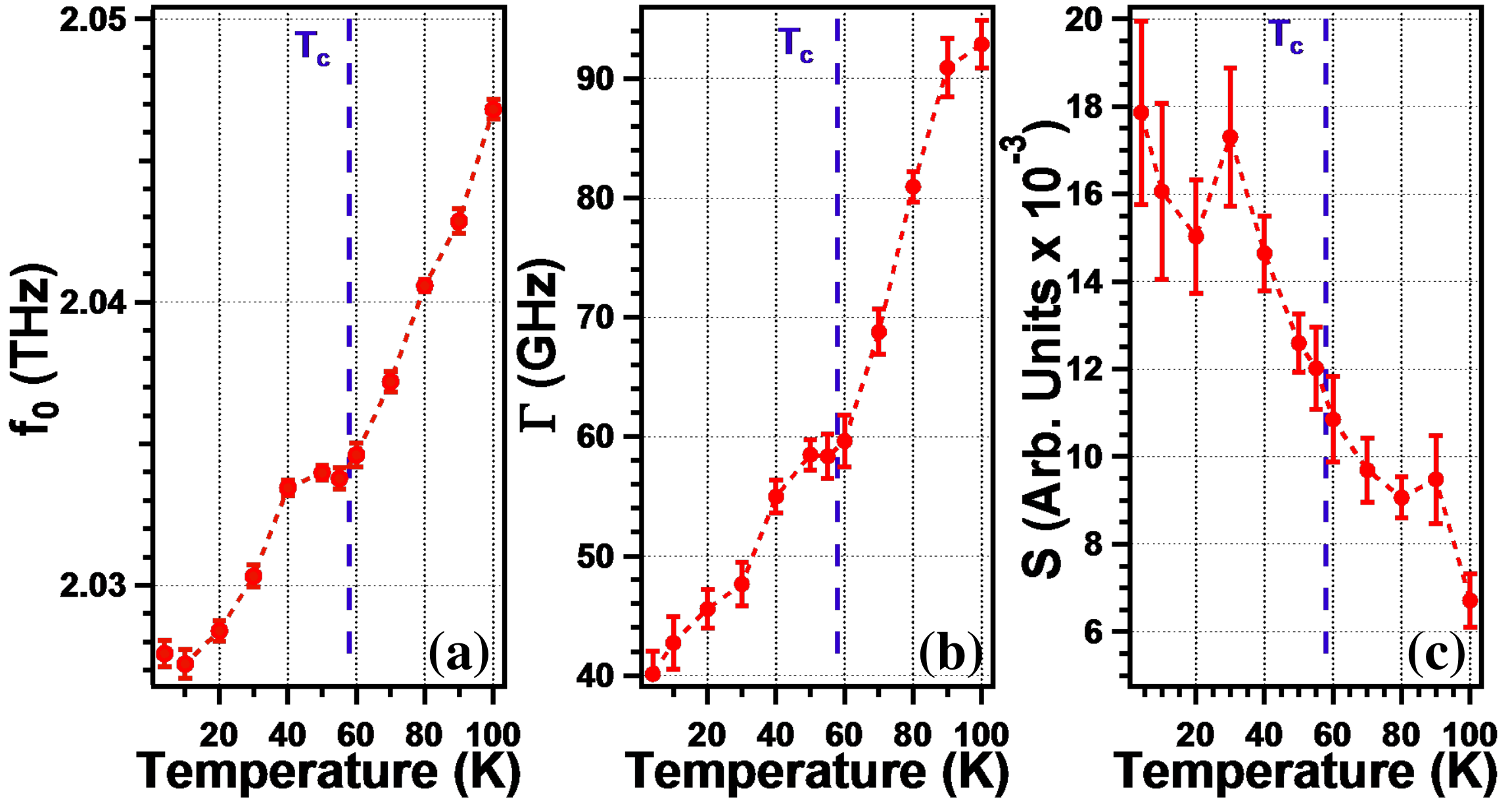}
\caption{Temperature dependent oscillator parameters, (a) frequency $f_0 = \omega _0 / 2 \pi$, (b) full width at half max $\Gamma$, and (c) oscillator strength $S$, of the magnon shown in \autoref{Fig2}.  Error bars are based on the quality of the fits.  A clear anomaly in the magnon's frequency and width can be seen at T$_c$ $\approx$ 58K (vertical dashed lines), indicating a sensitivity to the magnetic transition.}
\label{Fig3}
\end{figure}

\autoref{Fig3} displays the temperature dependent oscillator parameters, (a) $f_0 = \omega _0 / 2 \pi$, (b) $\Gamma$, and (c) $S$, of the low frequency magnon as determined from fitting the spectra with Eq. \ref{Eq1}.  An additional linear background was included in the fits to account for the high frequency phonon at 2.5 THz.  Error bars in the figure are based on the quality of the fits. Unlike the results reported from Miller \textit{et al.} \cite{Miller2010}, we uncover a coupling of this excitation to the magnetic structure of Cu$_2$OSeO$_3$.  One can see in \autoref{Fig3}(a) that the magnon frequency displays a weak softening as the temperature is lowered, reducing by $\approx$ 1\% from 100K to 4K.  A clear anomaly is observed in the magnon's frequency at T$_c$ $\approx$ 58K (vertical dashed lines in \autoref{Fig3}), further supporting the magnetic character of this excitation.  A similar anomaly at T$_c$ is observed in the  width of the excitation, shown in \autoref{Fig3}(b), which shows the general trend of lower damping at lower temperatures.  The far infrared FTIR transmission spectroscopy experiments of Miller \textit{et al.} \cite{Miller2010} likely did not posses the level of sensitivity needed to observe these features which explains why this softening and sensitivity to the magnetic structure was not previously observed.  However, our experiments are able to determine the magnon frequency to a precision of $\approx$ 0.5 GHz (2 $\mu$eV), allowing for detection of such subtle effects.  

It may seem unusual that this magnetic excitation persists to such high temperatures, well above the magnetic transition at T$_c$.  However, this mode physically corresponds to the rigid rotation of all the spins of a single tetrahedra \cite{Romhanyi2014}.   As mentioned above, these tetrahedra remain well defined entities far above T$_c$ due to the strong exchange interactions between spins within the tetrahedra.  Therefore, the observation of this mode up to 120 K is consistent with the strongly entangled tetrahedra picture of Cu$_2$OSeO$_3$ \cite{Janson2014}.  Interestingly, we find this excitation does not display any discernible magnetic field dependence up to $\mu_0$H = 7 T in Faraday geometry.  We discuss this lack of field dependence, the folding of this zone boundary mode to the zone center, and the additional new observation of a gap in the magnon spectrum at the zone boundary in the discussion below.

\subsection{Magnetic Field Dependence}

\subsubsection{Magnetization Dependent Faraday and Kerr Rotations}

Rotation of the plane of polarization of incident radiation upon transmission (Faraday rotation) or reflection (Kerr rotation) can often be related to the underlying symmetry of the material under investigation.  For instance, the non-centrosymmetric chiral structure of Cu$_2$OSeO$_3$ permits natural optical activity, rotation of the plane of polarization of linearly polarized light upon transmission in zero applied magnetic field, an effect which was recently observed in the visible range \cite{Versteeg2016}.  Additional gyrotropic effects can occur when time reversal symmetry is broken, for instance by the spontaneous magnetization of the sample.  In this case the index of refraction matrix is fully antisymmetric with off-diagonal terms, assuming linear response, proportional to the sample magnetization \cite{Freiser1968}.  A circular basis transformation reveals that linearly polarized light undergoes Faraday and Kerr rotations proportional to the sample's magnetization upon transmission \cite{Woodford2007, Glazer2006, Barron2009, Freiser1968}.  Examination of the proportionality constants reveals that the Kerr rotation is expected to be weaker than the Faraday rotation by a factor of $\approx d / \lambda$, \cite{Freiser1968} which in the case of this experiment is $\approx$ 3.  Such magneto-optical effects allows one to treat polarization rotations as measures of the order parameter of the magnetically ordered phases and can therefore be used to construct a magnetic phase diagram.  

\begin{figure*}
\includegraphics[width=2.0\columnwidth, keepaspectratio]{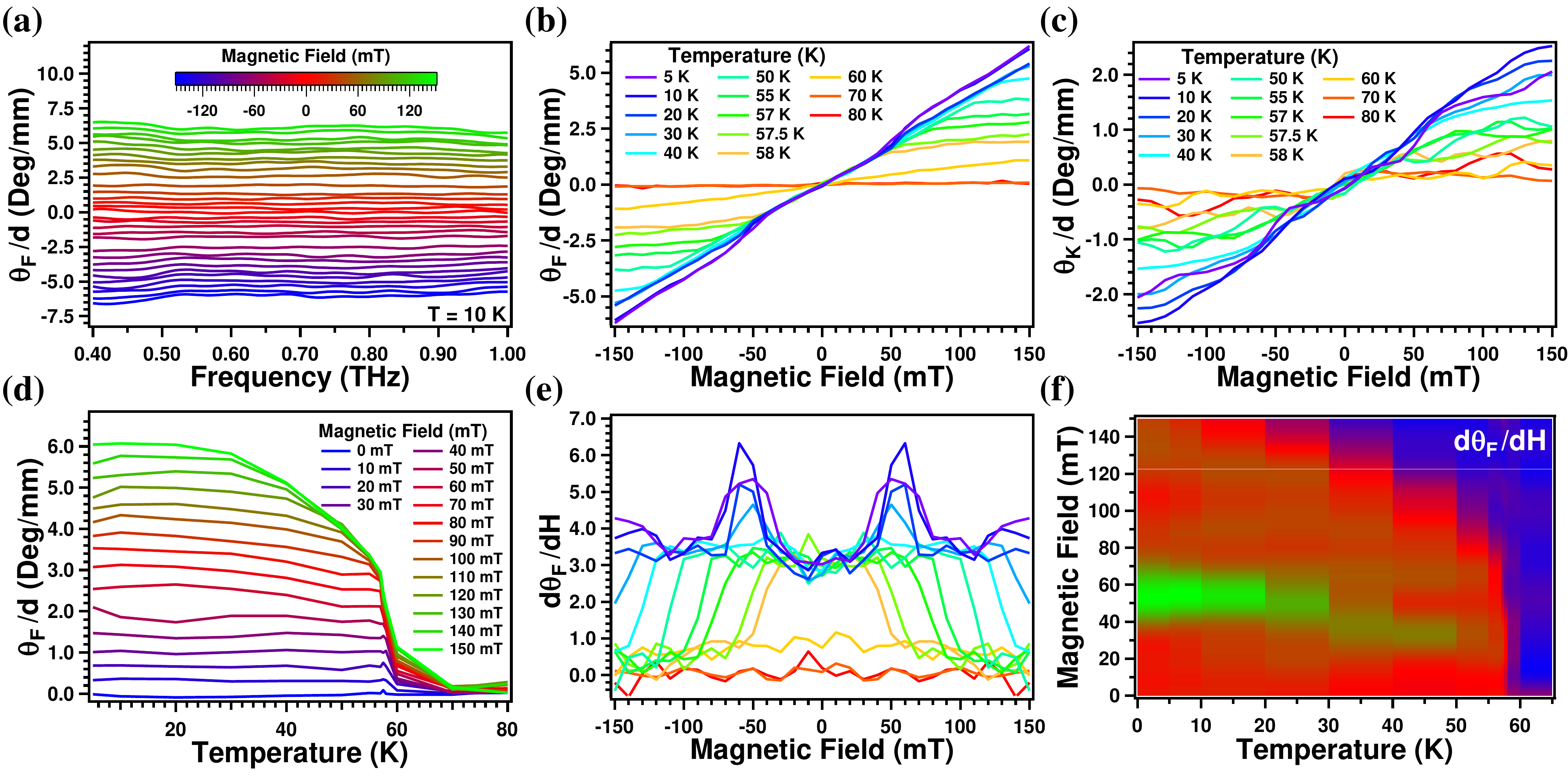}
\caption{(a) Frequency dependence of the real part of the Faraday rotation normalized by sample thickness at T = 10K.  (b-c) Real part of the (b) Faraday and (c) Kerr rotation as a function of magnetic field obtained by averaging the data over the frequency range shown in (a).  The proportionality between both the Faraday and Kerr rotations and the sample magnetization is clearly observed.  (d) Temperature dependence of the Faraday rotation.  (e) The field derivative of the Faraday rotation as a function of magnetic field, a quantity proportional to the magnetic susceptibility of the sample.  (f) Image plot of the data shown in (e) where the phase boundaries between the helical and conical phases (bright green) and conical and field polarized phases (red to blue) are clearly seen.}
\label{Fig4}
\end{figure*}

For a single pass transmission experiment we can write the total polarization rotation of Cu$_2$OSeO$_3$ as, 
\begin{equation}
\theta _{Tot} = \theta_{\text{NOA}} + \theta_F (M(H,T))
\end{equation}
where $\theta_{\text{NOA}}$ is the natural optical activity intrinsic to the chiral lattice of Cu$_2$OSeO$_3$, $\theta_F (M(H,T))$ is the complex magnetization dependent Faraday rotation, and higher order terms have been neglected.  The natural optical activity is too weak to observe in our long wavelength THz measurements as it scales inversely with the wavelength of light.  Instead, improved signal to noise is obtained by subtracting the zero field rotation from the field dependent data.  This is justified as although Cu$_2$OSeO$_3$ orders in zero magnetic field, the helical phase is marked by domain formation such that the net magnetization in this phase is zero.  If we define $\theta ^{'}_{\text{Tot}}(H,T)$ = $\theta_\text{Tot}(H,T) - \theta_{\text{Tot}}(H=0, T)$, then the field dependent polarization rotation, normalized by the sample thickness $d$, is given by,
\begin{equation}
\frac{1}{d} \theta ^{'}_{\text{Tot}}(H,T) =  \frac{1}{d} \theta _{F}(M(H, T))
\label{Fara}
\end{equation}

Additional information and enhanced signal to noise can be achieved by examining multiple reflections (``echos") of the THz pulse through the sample.  The symmetry and finite magnetization of Cu$_2$OSeO$_3$ results in a Faraday rotation that further rotates upon reflection inside the sample.  Therefore, the first echo of light, which travels through the sample a total of three times, gains a contribution to its rotation that is three times the Faraday rotation of the first transmitted pulse.  Additionally, the first echo also reflects internally off the sample surface twice, each time gaining a Kerr rotation that will also be magnetization dependent, but is expected to rotate in the opposite direction of the Faraday rotation \cite{Freiser1968}.  Therefore the total polarization rotation of the first reflected pulse is given by,
\begin{equation}
\frac{1}{d} \theta ^{'}_{\text{Tot}}(H,T) =  \frac{1}{d}[ 3\theta _{F}(M(H,T)) - 2 \theta _{K}(M(H,T))].
\label{Kerr}
\end{equation}
\noindent where, the first and second terms represent the Faraday and Kerr rotations respectively.  Thus, the complex Faraday and Kerr rotation angles can be measured independently if both the first transmitted and first reflected pulses of terahertz light through the sample are measured.

\autoref{Fig4} displays the results of our polarimetry experiments of Cu$_2$OSeO$_3$.  \autoref{Fig4}(a) shows the real part of the extracted Faraday rotation per mm of sample thickness, as defined in Eq. \ref{Fara}, as a function of frequency and applied magnetic field at T=10K.  The Faraday rotation in our spectral range shows little frequency dependence.  However, structure can be found in the field dependence of the data.  \autoref{Fig4}(b) shows the real part of the Faraday rotation as a function of magnetic field obtained from averaging the data in \autoref{Fig4}(a) over the frequency range shown at each temperature.  

The proportionality between the Faraday rotation and magnetization is easily observed in \autoref{Fig4}(b).  The Faraday rotation is small at temperatures above T$_c$ $\approx$ 58K.  Below T$_c$, the system enters the multi-domain helical phase where magnetic order develops but with multiple domains resulting in no net magnetization.  Therefore, no additional Faraday rotation that results from magnetic ordering is expected at temperatures below T$_c$ in zero applied field.  Once a magnetic field is applied the spins cant in the direction of magnetic field resulting in a linear increase in magnetization and therefore an identical trend in Faraday rotation.  At H$_{c1}$ the helices co-align and the system enters the single-domain conical phase, which is accompanied by a jump in magnetization.  The corresponding increase in Faraday rotation can be observed for fields H$_{c1}$ $\approx$ $\pm$50 mT at T=5K.  At larger magnetic fields, H $\ge$ H$_{c2}$, the system enters the field polarized phase resulting in a saturation of the magnetization and Faraday rotation.  \autoref{Fig4}(c) displays the real part of the Kerr rotation as a function of magnetic field, obtained in a similar manner as the Faraday rotation described above and Eq. \ref{Kerr}.  The Kerr rotation displays an identical dependence on sample magnetization but is approximately a third that of the Faraday rotation, as expected from the ratio of d/$\lambda$.  

\begin{figure}
\includegraphics[width=1.0\columnwidth, height=2.0 in]{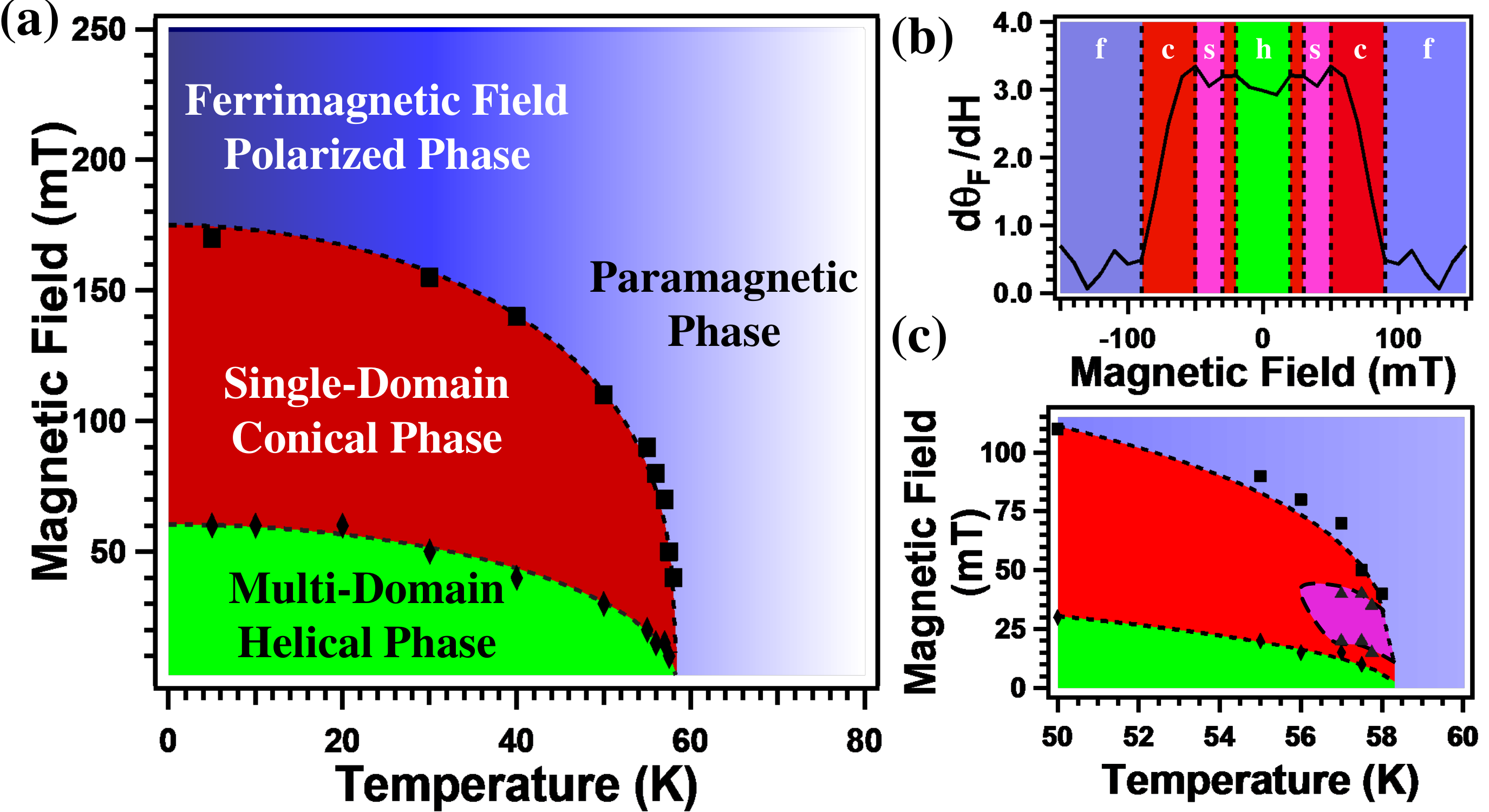}
\caption{(a) The H-T magnetic phase diagram of Cu$_2$OSeO$_3$ constructed from our Faraday rotation experiments shown in Figure  \ref{Fig4}.  (b) The first derivative of the Faraday rotation with respect to field at T=57K, a temperature at which all of the magnetic phases of Cu$_2$OSeO$_3$ can be observed.  Dashed vertical lines and distinct colors signify transitions from the (h)elical, (s)kyrmion, (c)onical, and (f)errimagnetic phases.  (c) Zoomed in region of the phase diagram where the skyrmion phase is observed.  Dotted phase boundary around the skyrmion phase is meant as a guide to the eye.   See text for more details.}
\label{Fig5}
\end{figure}

As the Faraday and Kerr rotations are proportional to the sample magnetization, an H-T phase diagram of Cu$_2$OSeO$_3$ can be constructed from the data shown in \autoref{Fig4} (a)-(c).  Here we focus primarily on the Faraday rotation, as the signal to noise is much better than that of the Kerr rotation due to technical aspects of our measurement.  In order to appropriately identify phase boundaries, subtle features in the data must be identified which are more easily observed in the temperature dependence and field derivatives of the Faraday rotation.  \autoref{Fig4}(d) displays the temperature dependence of the Faraday rotation at constant fields, where the phase boundary between the conical and field polarized phases is evident.  \autoref{Fig4}(e) shows the first derivative of the Faraday rotation with respect to magnetic field at constant temperatures, a quantity proportional to the magnetic susceptibility of the sample.  The transition from the helical to the conical state is now easily identified as a sharp maximum in the derivative.  The phase boundary between the conical and field polarized phases can be identified as the field beyond which the first derivative is zero or identically as a sharp maximum in the second derivative. \autoref{Fig4}(f) shows an image plot of the data in \autoref{Fig4}(e) in which clear phase boundaries at H$_{c1}$ (bright green) and H$_{c2}$ (red to blue) are easily observed.   

Figures \ref{Fig5}(a) displays our extracted H-T phase diagram of Cu$_2$OSeO$_3$ as determined from our polarization rotation experiments.  Symbols are the extracted phase boundaries from the data shown in \autoref{Fig4} while dotted lines result from power law fits of the data given by the expression, $H_{c}(T) = H_{c}(0) {(1 - (T / T_c)^ \alpha)}^{\beta}$, which was previously found to describe the data in both $\mu$SR \cite{Maisurandze2011} and ac susceptibility  \cite{Zivkovic2012, Zivkovic2014} investigations.  We restrict the critical exponent $\alpha$=2 as has been done previously \cite{Maisurandze2011, Zivkovic2012, Zivkovic2014} and is expected for a three dimensional system \cite{Kobler1999}.  From these fits we extract a critical temperature of T$_c$ = 58.4$\pm$0.4 K and a critical exponent of $\beta$ = 0.35$\pm$0.04 at H$_{c2}$.  Our extracted critical temperature is in excellent agreement with previous investigations.  While our extracted value of $\beta$ at H$_{c2}$ is in reasonable agreement with the $\beta$ = 0.367 of the 3D Heisenberg model and the $\beta$ $\approx$ 0.37 - 0.39 found in previous experiments of Cu$_2$OSeO$_3$ \cite{Maisurandze2011, Zivkovic2012, Zivkovic2014}.  

\autoref{Fig5}(b) displays the derivative of the Faraday rotation with respect to field at T=57K, a temperature at which all of the magnetic phases of Cu$_2$OSeO$_3$ can be observed.  Dotted lines and distinct colors mark different magnetic phases.  The skyrmion phase manifests in the magnetic susceptibility, and therefore in the derivative of the Faraday rotation, as an additional minimum shown in pink in \autoref{Fig5}(b).  \autoref{Fig5}(c) displays the phase diagram in the vicinity of T$_c$ in which the skyrmion phase can observed.  Although our data possesses limited temperature resolution and demagnetization effects of the sample have not been taken into account, the extracted phase diagram, including the skyrmion phase, is in excellent agreement with those reported in previous studies \cite{Adams2012, SekiA2012}.  

\subsubsection{THz Dynamics of the Uniform Mode}

While one can obtain an approximate understanding of the magnetic phases of Cu$_2$OSeO$_3$ by reducing the unit cell to four weakly coupled effective S=1 spins, an understanding of the excitation spectrum requires consideration of the full spin Hamiltonian in conjunction with quantum fluctuations.  Such a full quantum treatment has been performed by Janson \textit{et al.} \cite{Janson2014} and Romh\'anyi \textit{et al.} \cite{Romhanyi2014}, while corresponding neutron scattering \cite{Portnichenko2016, Tucker2016}, high field THz ESR \cite{Ozerov2014}, and Raman spectroscopy \cite{Gnezdilov2010} experiments reveal a striking agreement between the theoretical and experimentally observed excitation spectrums. 

Of particular importance to this work is the lowest energy excitation of Cu$_2$OSeO$_3$.  At the single tetrahedron level, the ground state is a three-fold degenerate triplet comprised of states with quantum numbers $\ket{S, S_z} = \ket{1, -1}$, $\ket{1,0}$, $\ket{1,1}$.  Each of these states are themselves a coherent quantum superposition of four classical ground states \cite{Janson2014, Romhanyi2014}.  Turning on interactions between tetrahedra at the mean field level mixes single tetrahedron states with identical symmetry.  The new resultant ground state is then a non-degenerate superposition of the original $\ket{1,1}$ triplet state and a higher energy $\ket{2,1}$ quintet state with wavefunction $\ket{\psi}_t$ = $\cos{(\alpha /2)}\ket{1,1} + \sin{(\alpha/2)}\ket{2,1}$, where the variational parameter $\alpha$ controls the degree of mixing \cite{Janson2014, Romhanyi2014}.  One can see that the ground state wavefunction of each tetrahedron is no longer a state of definite angular momentum and states can only be labeled by their $S_z$ components.  Including quantum fluctuations into the theory renormalizes the excitation spectrum.  In zero field the lowest energy excitation is a parabolically dispersing Goldstone mode associated with the reduction of symmetry from SU(2) to U(1) in the ferrimagnetic state \cite{Portnichenko2016, Tucker2016, Janson2014, Romhanyi2014}.  Application of magnetic field gaps the Goldstone mode, which is hereafter referred to as the uniform mode, by an amount proportional to the field through Zeeman coupling \cite{Ozerov2014, Kobets2010}.     

\begin{figure}[h]
\includegraphics[width=0.95\columnwidth, keepaspectratio]{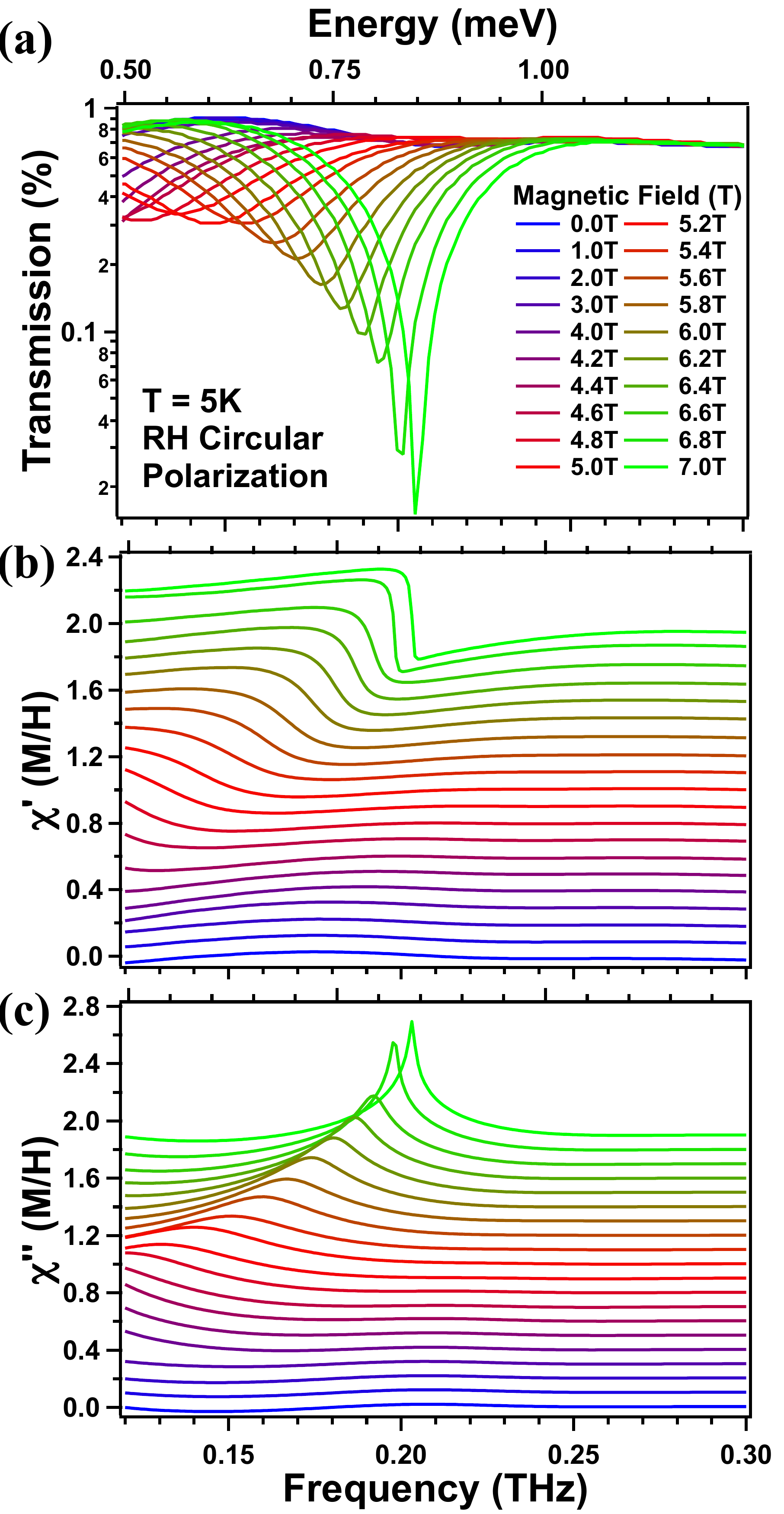}
\caption{Results of our high field transmission experiments of Cu$_2$OSeO$_3$ presented in the circular basis with only the right hand channel shown.  (a) Magnitude of the complex transmission as a function of magnetic field and frequency at T=5K.  A sharp magnetic absorption is observed at low frequencies which we identify as the uniform mode of the field polarized phase.  (b)-(c) Corresponding (b) real $\chi ^{'}(\omega)$ and (c) imaginary $\chi ^{''}(\omega)$ parts of the complex magnetic susceptibility. Offsets of 0.1 per field have been added for clarity.}
\label{Fig6}
\end{figure}

\autoref{Fig6} displays the results of our high field transmission experiments in which the uniform mode is observed.  Improved signal to noise and systematics were obtained by applying a cosine window function to the data in the time-domain before Fourier transforming.  Here data is presented in the right hand channel of the circular basis which, as discussed in the methods section above, is an eigenpolarization of the system.  We find that the uniform mode is only active to right hand circularly polarized light, as expected for a magnetic excitation with a well defined magnetic dipole moment of $\Delta S_z$=-1 \cite{Romhanyi2014}.  \autoref{Fig6}(a) displays the magnitude of the complex transmission of Cu$_2$OSeO$_3$ at T=5K as a function of magnetic field and frequency.   The uniform mode enters our accessible frequency range around $\mu_0$H $\approx$ 5T and can be seen to display an increase in resonant frequency and a narrowing width with increasing applied field.  \autoref{Fig6}(b)-(c) displays the (b) real and (c) imaginary parts of the magnetic susceptibility extracted from the data shown in \autoref{Fig6}(a) and Eq. \ref{Transeq}.  In order to extract the magnetic susceptibility, data were referenced to identical field scans at T=100K, a temperature at which the absorption is no longer observed.  The implicit assumption here is that the dielectric properties of Cu$_2$OSeO$_3$ do not appreciably change below 100K, typically a good assumption for such a large gap insulator \cite{Laurita2015}. 

\begin{figure}
\includegraphics[width=1\columnwidth, keepaspectratio ]{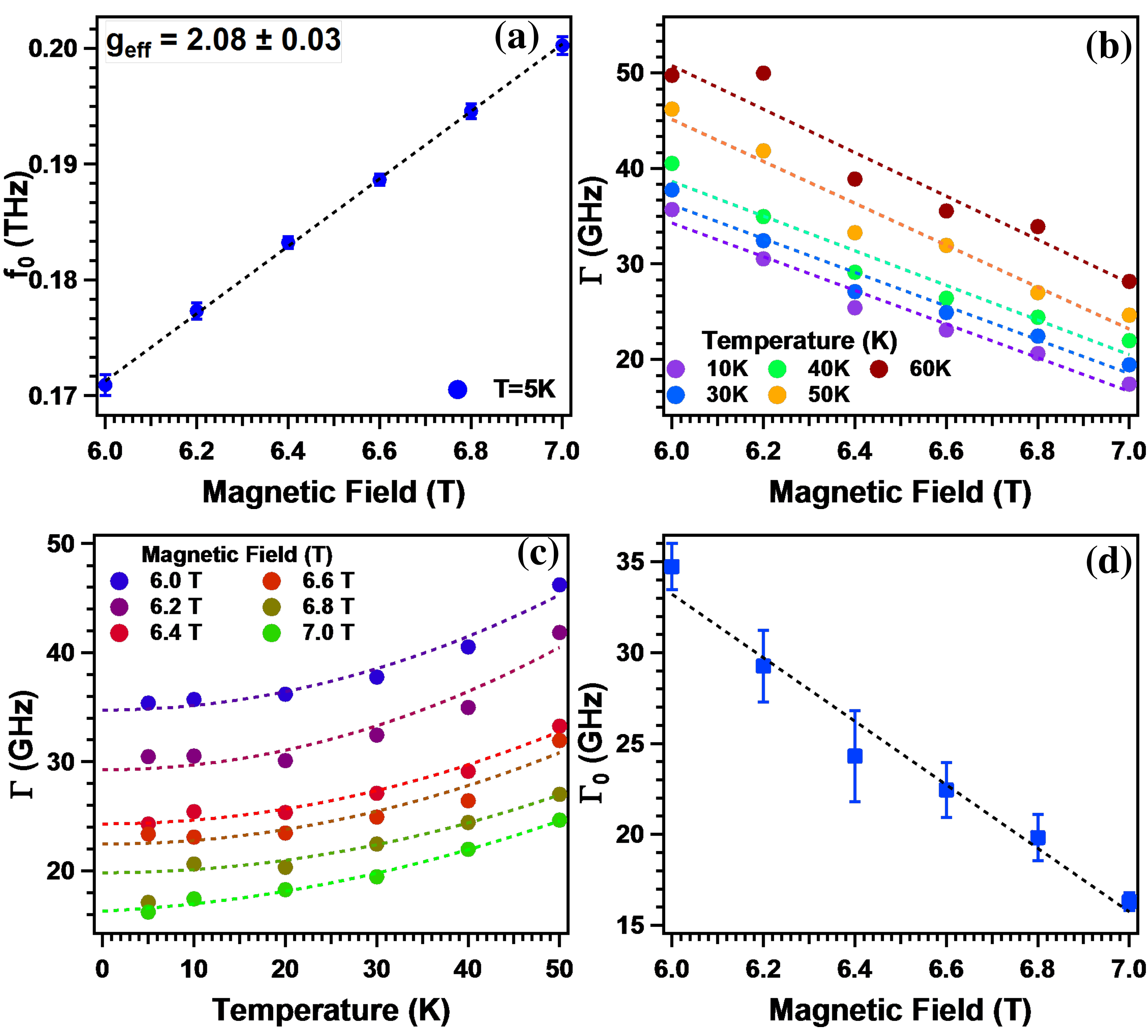}
\caption{Dynamical properties of the uniform mode obtained from fitting the susceptibility shown in \autoref{Fig6} to Eq. \ref{Eq1}.  (a) Field dependence of the resonant frequency at T=5K as well as a linear fit from which a $g_\text{eff}$ = 2.08 $\pm$ 0.03 is obtained. (b) Field dependence of the full width at half max ($\Gamma$) of the uniform mode at several representative temperatures. (c) Temperature dependence of the $\Gamma$ at several values of magnetic field.  Dashed lines are fits of the data by Eq. \ref{Decayeq} which reveals a zero temperature spontaneous decay rate.  (d) Magnetic field dependence of the spontaneous decay rate obtained from fitting the data shown in (c).  The dashed linear fit of the data is meant as a guide to the eye.}
\label{Fig7}
\end{figure}

The data presented in \autoref{Fig6} were fit to the general model given in Eq. \ref{Eq1} in order to extract the dynamical properties of the uniform mode.  \autoref{Fig7}(a) displays the extracted resonant frequencies, $f_0=\omega _0 / 2 \pi$, as a function of magnetic field at T=5K, the temperature at which the highest resolution of our measurement is obtained.  Error bars are based on the quality of the fits.  The dotted line is a linear fit of the data as expected for Zeeman coupling.  From this fit we obtain an effective $g$-factor of $g_\text{eff}$ = 2.08 $\pm$ 0.03, which is in excellent agreement with the expected value for Cu$^{2+}$ spins and the THz ESR measurements of Ozerov \textit{et al.} \cite{Ozerov2014} which previously reported $g_\text{eff}$ = 2.1 $\pm$ 0.1.  

Additional information regarding the dynamics of this mode can be obtained by examining the width of the excitation as a function of temperature and magnetic field.  In the limit of no disorder, the excitation width represents the decay rate ($\Gamma$), or the inverse lifetime, of the uniform mode.  \autoref{Fig7}(b) displays the field dependence of the width of the uniform mode at several temperatures.  The width displays an unusual approximately linear decrease in the accessible field region of our measurement for all temperatures, suggesting a dominant non-Gilbert damping mechanism.  The temperature dependence of the width of the uniform mode at fields between 6T and 7T is shown in \autoref{Fig7}(c).  A broadening of the uniform mode with increasing temperature is observed.  Such thermal broadening can be ascribed to enhanced decay through interactions with thermally excited magnons, processes which become frozen out at low temperatures.  

The functional dependence of the decay rate with temperature may reveal additional information regarding the decay processes of the uniform mode.  Magnon decay through magnon-magnon interactions is a well studied topic dating back to the earliest days of spin wave theory \cite{Sparks1961, Schlomann1961, Kittel1953}.  In the simplest case, the spin wave Hamiltonian is completely harmonic, i.e. spin waves are non-interacting plane waves.  Interactions can be included by introducing anharmonic terms into the Hamiltonian which couple magnon states.  In general, such interaction terms do not conserve quasiparticle number and one must rely on symmetry and conservation laws to determine which decay channels are permitted \cite{Zhitomirsky2013}.  In the simplest cases, the temperature dependence of such decay processes can expressed as a polynomial expansion in temperature with terms proportional to T and T$^2$ for the lowest order three and four magnon interactions respectively \cite{Schlomann1961}.  Terms proportional to T$^3$ or greater result from higher order magnon-magnon interactions that are neglected in our analysis.  Far less conventional are magnon decays at zero temperature, i.e. spontaneous decays, which arise from quantum, not thermal, fluctuations \cite{Zhitomirsky2013}.  The spontaneous decay rate will in general be a function of magnetic field stemming from field dependence of the kinematic requirements \cite{Zhitomirsky2013, Chernyshev2012} which must be satisfied for decays to occur.

Therefore we can write the total decay rate of the uniform mode as a function of both field and temperature as 
\begin{equation}
\Gamma(T,H) = \Gamma_0(0,H) + A(H)T + B(H)T^2
\label{Decayeq}
\end{equation}
\noindent where the $\Gamma_0(0,H)$ is the spontaneous decay rate and the terms proportional to T and T$^2$ result from three and four magnon interactions respectively as described above.  

Dashed lines in \autoref{Fig7}(c) are fits of the data to Eq. \ref{Decayeq}.  The decay rate is well described by Eq. \ref{Decayeq} and an extrapolation of the fits to the zero temperature limit reveals a finite spontaneous decay rate.  \autoref{Fig7}(d) displays the field dependence of the extracted spontaneous decay rate obtained from the fits shown in \autoref{Fig7}(c).  One can observe that the  spontaneous decay rate displays an approximately negative linear dependence with magnetic field in the accessible region of our measurement.  The decay processes of the uniform mode and the possible origins of the spontaneous decay rate are addressed below.  

\section{Discussion}

Dzyaloshinkii-Moriya (DM) interactions are obviously vital in the low energy description of chiral magnets.  Yet, spin-wave calculations of Cu$_2$OSeO$_3$ \cite{Romhanyi2014, Portnichenko2016} have thus far not included DM interactions, effectively treating the system as a ferromagnet.  However, DM interactions in Cu$_2$OSeO$_3$ have been suggested to be exceptionally strong.  Recent calculations predict the largest DM interaction, referred to as $D_4$, to range from $|D_{4} / J_{4}|$ $\approx$ 0.5 \cite{Janson2014} to $|D_{4} / J_{4}|$ $\approx$ 1.95 \cite{Yang2012}, \textit{nearly twice the symmetric exchange}.  Thus, it is reasonable to assume that DM interactions may have a more profound impact in Cu$_2$OSeO$_3$ than other chiral magnets.  As it is expected that DM interactions modify the spin-wave spectrum at low energies near the $\Gamma$ point \cite{Portnichenko2016, Romhanyi2014}, the exact region probed by low energy optical spectroscopy, we believe that many of the observations made is this work can be attributed to such DM interactions.

In zero field, we observed a magnetic excitation with frequency  $f_0 \approx $ 2.03 THz (8.40 meV) which we demonstrated is a zone folded magnon from the zone boundary to the zone center.  It should be noted that Cu$_2$OSeO$_3$ displays no change in either structural or magnetic symmetry from T$_c$ down to at least 10K \cite{Bos2008}, suggesting a different mechanism for this folding.  Assuming spin-wave calculations, which treat the unit cell as an FCC lattice, have captured the symmetry of Cu$_2$OSeO$_3$ correctly, then we attribute this new magnon excitation to DM interactions which thus far have not been included in calculations.  We speculate that this mode is permitted by symmetry to exist at the $\Gamma$ point but is silent in the spin-wave calculations due to vanishing intensity.  However, DM interactions, which will presumably mix magnon states, may give intensity to this otherwise silent mode.  We hope this study will motivate future spin-wave calculations which include the DM interactions to further investigate. 

The lack of discernible field dependence of this mode may be consistent with the mechanism described above.  Spin-wave calculations predict a degeneracy between two magnon bands at the $\mathrm{R}$ point, the higher energy magnon being a singlet associated with rotating the spins of a single tetrahedra against the mean field exerted by neighboring tetrahedra \cite{Romhanyi2014}.  Presumably DM interactions will mix magnon states at this high degeneracy point, opening a gap in the magnetic spectrum.  One can see in \autoref{THzNeutron} that indeed a clear gap is observed at the zone edge at 9 meV, which is reported here for the first time.  We speculate that this mixing results in a band character at the extrema that is predominantly singlet, explaining the lack of field dependence when this mode is then folded to the zone center.  Again, spin-wave calculations which include DM interactions or neutron scattering  measurements in magnetic field, which have not yet been performed, would be needed to investigate this further.  

The helical, conical, and skyrmion phases of Cu$_2$OSeO$_3$ are stabilized by the competition between DM and Heisenberg exchange interactions.  In this work we showed that such phases can be detected by high resolution polarimetry experiments.  Here we only remark that it is surprising that the observed Faraday rotation in the THz range possesses no frequency dependence.  One would generally expect that the THz spectra would display signatures of the low frequency excitations of Cu$_2$OSeO$_3$, for instance the helimagnon skyrmion \cite{Schwarze2015} excitations or the uniform mode.  Although these excitations lie at lower frequencies than those probed by our measurements in zero or  small magnetic fields, spectral signatures of these excitations are generally expected to extend to higher frequencies.  Instead we observe Faraday and Kerr rotations with no discernible frequency dependence within our spectral range.

In large magnetic fields, we studied the field and temperature dependent dynamics of the uniform mode of Cu$_2$OSeO$_3$.  We found this excitation to surprisingly narrow with increasing applied magnetic field, suggesting a dominant non-Gilbert damping mechanism.  Such a narrowing with magnetic field is typically only observed in these chiral magnets in weak magnetic fields before entering the field polarized phase \cite{Schwarze2015}.  The origin of this narrowing is currently unclear.  However, there have been predictions of an additional weak antiferromagnet order that exists on top of the ferrimagnetic order in Cu$_2$OSeO$_3$ \cite{Janson2014}, and in MnSi type crystals in general \cite{Chizikov2015}, which results from an additional spin canting that persists into the field polarized phase.  We speculate that the narrowing of this excitation in field may stem from overcoming this canting in large fields, which would presumably reduce magnon coupling.  We hope our measurements will inspire future investigations into this effect.    

We also discovered that the uniform mode of Cu$_2$OSeO$_3$ possessed a spontaneous decay rate in the zero temperature limit.  One may be quick to ascribe such a zero temperature decay to inhomogeneous broadening from disorder.  While we cannot definitely rule out this possibility, the strong field dependence of the spontaneous decay rate may be indicative of a different origin.  In fact, there are several reasons to suspect that such spontaneous decays are permitted in Cu$_2$OSeO$_3$.  Spontaneous decays, which require anharmonic magnon interactions are, generally speaking, only permitted if two criteria are met \cite{Zhitomirsky2013}.  First, the spin order must be non-collinear due to symmetry and angular momentum conservation \cite{Zhitomirsky2013, Oguchi1960}.  Second, the magnon spectrum must be able to support magnon decays in a fashion that conserves both energy and momentum.  We address these points below.  

DM interactions are a natural mechanism to obtain a non-collinear spin structure and a coupling of transverse and longitudinal spin components, which may therefore permit anharmonic magnon interactions.  Additionally, as we mentioned above, the uniform mode of Cu$_2$OSeO$_3$ is not a state with well defined angular momentum but is instead a superposition of several angular momentum states.  One would generally expect that the quantum entangled nature of this state would result in zero point motion and therefore may lead to spontaneous decays.  Finally, the proposed additional antiferromagnetic order described above would likely also couple magnon states in such a fashion to result in a spontaneous decay.  Further theoretical and experimental research is needed to investigate if these effects, or perhaps others, can account for the observed spontaneous decay rate.

With such anharmonic terms possibly allowed the question remains how the uniform mode, the expected global minimum of the spin wave spectrum, can decay while conserving energy.  While the minimum of the magnon band is expected to be the uniform mode at $\vec{k}$=0 from spin wave theory, weak dipolar interactions, which are always present in ferromagnets, raise the energy of the uniform mode in magnetic field by an amount proportional to the sample magnetization \cite{Rezende2009, Chernyshev2012, Ordonez2009, Kreisel2009}.  The resulting band structure then contains minima at small momenta $\vec{k}_\text{min}$ $>$ 0, the exact value of  which depends on sample geometry.  Therefore, the uniform mode can then in principle decay by splitting into magnons at the band minima assuming the kinematic requirements are met.  Such magnon splitting through dipolar effects have been extensively studied in the similar compound YIG \cite{Chernyshev2012, Ordonez2009, Rezende2009, Kreisel2009}.  Theoretical treatments which include dipolar effects are needed to determine if such effects can account for the observed decay of the uniform mode in Cu$_2$OSeO$_3$.  

We also note that it is also generally possible to observe a broadening of a resonance peak by non-equilibrium effects, in the form of a four magnon anharmonic interaction \cite{Wigen1994}.  Such effects have been observed in microwave resonance experiments.  However, the fields used in our THz measurements are substantially weaker than those of typical microwave resonance experiments and it is generally assumed that our experiments are strictly in the linear response regime.  Therefore, we remark that while the spontaneous decay of the uniform mode may be caused by quantum interactions, further measurements and investigations are required to fully understand the origin of the spontaneous decay.

\section{Conclusion}

In this work, high resolution terahertz transmission and polarimetry experiments were utilized to probe the magneto-optics of the skyrmion insulator Cu$_2$OSeO$_3$.  Experiments performed throughout the magnetic phase diagram uncovered a new magnetic excitation which was shown to be folded from the zone boundary to the zone center, detected the magnetic phases including the skyrmion phase, and unveiled the unusual dynamics of the uniform precession of the field polarized phase.  These observations were generally attributed to the effects of DM interactions, which may be particularly strong in Cu$_2$OSeO$_3$ and are generally expected to impact the low energy magnetic response of this chiral magnet.  Our results underline the need for further investigation into the effects of DM interactions in these systems.

\section{Acknowledgments}

This research was funded by the U.S. Department of Energy, Office of Basic Energy Sciences, Division of Materials Sciences and Engineering through Grant No. DE-FG02-08ER46544. NJL acknowledges additional support through the ARCS Foundation Dillon Fellowship.  GGM acknowledges generous support from the NSF-GRFP, Grant No. DGE-1232825.  We would like to thank L. Balents, S. Chernyshev, W. Fuhrman, F, Mahmood, K. Plumb, M. Valentine, C. Varma for helpful conversations. 

\bibliography{CuOSeOBib}

\begin{thebibliography}{75}
\expandafter\ifx\csname natexlab\endcsname\relax\def\natexlab#1{#1}\fi
\expandafter\ifx\csname bibnamefont\endcsname\relax
  \def\bibnamefont#1{#1}\fi
\expandafter\ifx\csname bibfnamefont\endcsname\relax
  \def\bibfnamefont#1{#1}\fi
\expandafter\ifx\csname citenamefont\endcsname\relax
  \def\citenamefont#1{#1}\fi
\expandafter\ifx\csname url\endcsname\relax
  \def\url#1{\texttt{#1}}\fi
\expandafter\ifx\csname urlprefix\endcsname\relax\def\urlprefix{URL }\fi
\providecommand{\bibinfo}[2]{#2}
\providecommand{\eprint}[2][]{\url{#2}}

\bibitem[{\citenamefont{Skyrme}(1962)}]{Skyrme1962}
\bibinfo{author}{\bibfnamefont{T.~H.~R.} \bibnamefont{Skyrme}},
  \bibinfo{journal}{Nucl. Phys.} \textbf{\bibinfo{volume}{31}}
  (\bibinfo{year}{1962}).

\bibitem[{\citenamefont{Nagaosa and Tokura}(2013)}]{Nagaosa2013}
\bibinfo{author}{\bibfnamefont{N.}~\bibnamefont{Nagaosa}} \bibnamefont{and}
  \bibinfo{author}{\bibfnamefont{Y.}~\bibnamefont{Tokura}},
  \bibinfo{journal}{Nature Nanotechnology} \textbf{\bibinfo{volume}{8}},
  \bibinfo{pages}{899} (\bibinfo{year}{2013}).

\bibitem[{\citenamefont{Bogdanov and Yablonskii}(1989)}]{Bogdanov1989}
\bibinfo{author}{\bibfnamefont{A.~N.} \bibnamefont{Bogdanov}} \bibnamefont{and}
  \bibinfo{author}{\bibfnamefont{D.~A.} \bibnamefont{Yablonskii}},
  \bibinfo{journal}{Sov. Phys. JETP} \textbf{\bibinfo{volume}{68}}
  (\bibinfo{year}{1989}).

\bibitem[{\citenamefont{Rozler et~al.}(2006)\citenamefont{Rozler, Bogdanov, and
  Pfleiderer}}]{Rozler2006}
\bibinfo{author}{\bibfnamefont{U.~K.} \bibnamefont{Rozler}},
  \bibinfo{author}{\bibfnamefont{A.~N.} \bibnamefont{Bogdanov}},
  \bibnamefont{and}
  \bibinfo{author}{\bibfnamefont{C.}~\bibnamefont{Pfleiderer}},
  \bibinfo{journal}{Nature} \textbf{\bibinfo{volume}{442}},
  \bibinfo{pages}{797} (\bibinfo{year}{2006}).

\bibitem[{\citenamefont{M\"uhlbauer et~al.}(2009)\citenamefont{M\"uhlbauer,
  Binz, Jonietz, Pfleiderer, Rosch, Neubauer, Georgii, and
  Boni}}]{Muhlbauer2009}
\bibinfo{author}{\bibfnamefont{S.}~\bibnamefont{M\"uhlbauer}},
  \bibinfo{author}{\bibfnamefont{B.}~\bibnamefont{Binz}},
  \bibinfo{author}{\bibfnamefont{F.}~\bibnamefont{Jonietz}},
  \bibinfo{author}{\bibfnamefont{C.}~\bibnamefont{Pfleiderer}},
  \bibinfo{author}{\bibfnamefont{A.}~\bibnamefont{Rosch}},
  \bibinfo{author}{\bibfnamefont{A.}~\bibnamefont{Neubauer}},
  \bibinfo{author}{\bibfnamefont{R.}~\bibnamefont{Georgii}}, \bibnamefont{and}
  \bibinfo{author}{\bibfnamefont{P.}~\bibnamefont{Boni}},
  \bibinfo{journal}{Science} \textbf{\bibinfo{volume}{323}},
  \bibinfo{pages}{915} (\bibinfo{year}{2009}).

\bibitem[{\citenamefont{Yu et~al.}(2011)\citenamefont{Yu, Kanazawa, Onose,
  Kimoto, Zhang, Matsui, and Tokura}}]{Yu2011}
\bibinfo{author}{\bibfnamefont{X.~Z.} \bibnamefont{Yu}},
  \bibinfo{author}{\bibfnamefont{J.}~\bibnamefont{Kanazawa}},
  \bibinfo{author}{\bibfnamefont{Y.}~\bibnamefont{Onose}},
  \bibinfo{author}{\bibfnamefont{K.}~\bibnamefont{Kimoto}},
  \bibinfo{author}{\bibfnamefont{W.~Z.} \bibnamefont{Zhang}},
  \bibinfo{author}{\bibfnamefont{Y.}~\bibnamefont{Matsui}}, \bibnamefont{and}
  \bibinfo{author}{\bibfnamefont{Y.}~\bibnamefont{Tokura}},
  \bibinfo{journal}{Nat. Mater.} \textbf{\bibinfo{volume}{10}},
  \bibinfo{pages}{106} (\bibinfo{year}{2011}).

\bibitem[{\citenamefont{Yu et~al.}(2010)\citenamefont{Yu, Onose, Kanazawa,
  Park, Han, Matsui, Nagaosa, and Tokura}}]{Yu2010}
\bibinfo{author}{\bibfnamefont{X.~Z.} \bibnamefont{Yu}},
  \bibinfo{author}{\bibfnamefont{Y.}~\bibnamefont{Onose}},
  \bibinfo{author}{\bibfnamefont{J.}~\bibnamefont{Kanazawa}},
  \bibinfo{author}{\bibfnamefont{J.~H.} \bibnamefont{Park}},
  \bibinfo{author}{\bibfnamefont{J.~H.} \bibnamefont{Han}},
  \bibinfo{author}{\bibfnamefont{Y.}~\bibnamefont{Matsui}},
  \bibinfo{author}{\bibfnamefont{N.}~\bibnamefont{Nagaosa}}, \bibnamefont{and}
  \bibinfo{author}{\bibfnamefont{Y.}~\bibnamefont{Tokura}},
  \bibinfo{journal}{Nature} \textbf{\bibinfo{volume}{465}},
  \bibinfo{pages}{901} (\bibinfo{year}{2010}).

\bibitem[{\citenamefont{M\"unzer et~al.}(2010)\citenamefont{M\"unzer, Neubauer,
  Adams, M\"uhlbauer, Franz, Jonietz, Georgii, B\"oni, Pedersen, Schmidt
  et~al.}}]{Munzer2010}
\bibinfo{author}{\bibfnamefont{W.}~\bibnamefont{M\"unzer}},
  \bibinfo{author}{\bibfnamefont{A.}~\bibnamefont{Neubauer}},
  \bibinfo{author}{\bibfnamefont{T.}~\bibnamefont{Adams}},
  \bibinfo{author}{\bibfnamefont{S.}~\bibnamefont{M\"uhlbauer}},
  \bibinfo{author}{\bibfnamefont{C.}~\bibnamefont{Franz}},
  \bibinfo{author}{\bibfnamefont{F.}~\bibnamefont{Jonietz}},
  \bibinfo{author}{\bibfnamefont{R.}~\bibnamefont{Georgii}},
  \bibinfo{author}{\bibfnamefont{P.}~\bibnamefont{B\"oni}},
  \bibinfo{author}{\bibfnamefont{B.}~\bibnamefont{Pedersen}},
  \bibinfo{author}{\bibfnamefont{M.}~\bibnamefont{Schmidt}},
  \bibnamefont{et~al.}, \bibinfo{journal}{Phys. Rev. B}
  \textbf{\bibinfo{volume}{81}}, \bibinfo{pages}{041203}
  (\bibinfo{year}{2010}).

\bibitem[{\citenamefont{Schulz et~al.}(2012)\citenamefont{Schulz, Ritz, Bauer,
  Halder, Wager, Franz, Pfleiderer, Everschor, Garst, and Rosch}}]{Schultz2012}
\bibinfo{author}{\bibfnamefont{T.}~\bibnamefont{Schulz}},
  \bibinfo{author}{\bibfnamefont{R.}~\bibnamefont{Ritz}},
  \bibinfo{author}{\bibfnamefont{A.}~\bibnamefont{Bauer}},
  \bibinfo{author}{\bibfnamefont{M.}~\bibnamefont{Halder}},
  \bibinfo{author}{\bibfnamefont{M.}~\bibnamefont{Wager}},
  \bibinfo{author}{\bibfnamefont{C.}~\bibnamefont{Franz}},
  \bibinfo{author}{\bibfnamefont{C.}~\bibnamefont{Pfleiderer}},
  \bibinfo{author}{\bibfnamefont{K.}~\bibnamefont{Everschor}},
  \bibinfo{author}{\bibfnamefont{M.}~\bibnamefont{Garst}}, \bibnamefont{and}
  \bibinfo{author}{\bibfnamefont{A.}~\bibnamefont{Rosch}},
  \bibinfo{journal}{Nat. Phys.} \textbf{\bibinfo{volume}{8}},
  \bibinfo{pages}{301} (\bibinfo{year}{2012}).

\bibitem[{\citenamefont{Jonietz et~al.}(2010)\citenamefont{Jonietz,
  M{\"u}hlbauer, Pfleiderer, Neubauer, M{\"u}nzer, Bauer, Adams, Georgii,
  B{\"o}ni, Duine et~al.}}]{Jonietz2010}
\bibinfo{author}{\bibfnamefont{F.}~\bibnamefont{Jonietz}},
  \bibinfo{author}{\bibfnamefont{S.}~\bibnamefont{M{\"u}hlbauer}},
  \bibinfo{author}{\bibfnamefont{C.}~\bibnamefont{Pfleiderer}},
  \bibinfo{author}{\bibfnamefont{A.}~\bibnamefont{Neubauer}},
  \bibinfo{author}{\bibfnamefont{W.}~\bibnamefont{M{\"u}nzer}},
  \bibinfo{author}{\bibfnamefont{A.}~\bibnamefont{Bauer}},
  \bibinfo{author}{\bibfnamefont{T.}~\bibnamefont{Adams}},
  \bibinfo{author}{\bibfnamefont{R.}~\bibnamefont{Georgii}},
  \bibinfo{author}{\bibfnamefont{P.}~\bibnamefont{B{\"o}ni}},
  \bibinfo{author}{\bibfnamefont{R.~A.} \bibnamefont{Duine}},
  \bibnamefont{et~al.}, \bibinfo{journal}{Science}
  \textbf{\bibinfo{volume}{330}}, \bibinfo{pages}{1648} (\bibinfo{year}{2010}).

\bibitem[{\citenamefont{Seki et~al.}(2012{\natexlab{a}})\citenamefont{Seki, Yu,
  Ishiwata, and Tokura}}]{SekiA2012}
\bibinfo{author}{\bibfnamefont{S.}~\bibnamefont{Seki}},
  \bibinfo{author}{\bibfnamefont{X.~Z.} \bibnamefont{Yu}},
  \bibinfo{author}{\bibfnamefont{S.}~\bibnamefont{Ishiwata}}, \bibnamefont{and}
  \bibinfo{author}{\bibfnamefont{Y.}~\bibnamefont{Tokura}},
  \bibinfo{journal}{Science} \textbf{\bibinfo{volume}{336}},
  \bibinfo{pages}{198} (\bibinfo{year}{2012}{\natexlab{a}}).

\bibitem[{\citenamefont{Seki et~al.}(2012{\natexlab{b}})\citenamefont{Seki,
  Kim, Inosov, Georgii, Keimer, Ishiwata, and Tokura}}]{SekiB2012}
\bibinfo{author}{\bibfnamefont{S.}~\bibnamefont{Seki}},
  \bibinfo{author}{\bibfnamefont{J.-H.} \bibnamefont{Kim}},
  \bibinfo{author}{\bibfnamefont{D.~S.} \bibnamefont{Inosov}},
  \bibinfo{author}{\bibfnamefont{R.}~\bibnamefont{Georgii}},
  \bibinfo{author}{\bibfnamefont{B.}~\bibnamefont{Keimer}},
  \bibinfo{author}{\bibfnamefont{S.}~\bibnamefont{Ishiwata}}, \bibnamefont{and}
  \bibinfo{author}{\bibfnamefont{Y.}~\bibnamefont{Tokura}},
  \bibinfo{journal}{Phys. Rev. B} \textbf{\bibinfo{volume}{85}},
  \bibinfo{pages}{220406} (\bibinfo{year}{2012}{\natexlab{b}}).

\bibitem[{\citenamefont{White et~al.}(2012)\citenamefont{White, Levatić,
  Omrani, Egetenmeyer, Prša, Živković, Gavilano, Kohlbrecher, Bartkowiak,
  Berger et~al.}}]{White2012}
\bibinfo{author}{\bibfnamefont{J.~S.} \bibnamefont{White}},
  \bibinfo{author}{\bibfnamefont{I.}~\bibnamefont{Levatić}},
  \bibinfo{author}{\bibfnamefont{A.~A.} \bibnamefont{Omrani}},
  \bibinfo{author}{\bibfnamefont{N.}~\bibnamefont{Egetenmeyer}},
  \bibinfo{author}{\bibfnamefont{K.}~\bibnamefont{Prša}},
  \bibinfo{author}{\bibfnamefont{I.}~\bibnamefont{Živković}},
  \bibinfo{author}{\bibfnamefont{J.~L.} \bibnamefont{Gavilano}},
  \bibinfo{author}{\bibfnamefont{J.}~\bibnamefont{Kohlbrecher}},
  \bibinfo{author}{\bibfnamefont{M.}~\bibnamefont{Bartkowiak}},
  \bibinfo{author}{\bibfnamefont{H.}~\bibnamefont{Berger}},
  \bibnamefont{et~al.}, \bibinfo{journal}{Journal of Physics: Condensed Matter}
  \textbf{\bibinfo{volume}{24}}, \bibinfo{pages}{432201}
  (\bibinfo{year}{2012}).

\bibitem[{\citenamefont{Yang et~al.}(2012)\citenamefont{Yang, Li, Lu, Whangbo,
  Wei, Gong, and Xiang}}]{Yang2012}
\bibinfo{author}{\bibfnamefont{J.~H.} \bibnamefont{Yang}},
  \bibinfo{author}{\bibfnamefont{Z.~L.} \bibnamefont{Li}},
  \bibinfo{author}{\bibfnamefont{X.~Z.} \bibnamefont{Lu}},
  \bibinfo{author}{\bibfnamefont{M.-H.} \bibnamefont{Whangbo}},
  \bibinfo{author}{\bibfnamefont{S.-H.} \bibnamefont{Wei}},
  \bibinfo{author}{\bibfnamefont{X.~G.} \bibnamefont{Gong}}, \bibnamefont{and}
  \bibinfo{author}{\bibfnamefont{H.~J.} \bibnamefont{Xiang}},
  \bibinfo{journal}{Phys. Rev. Lett.} \textbf{\bibinfo{volume}{109}},
  \bibinfo{pages}{107203} (\bibinfo{year}{2012}).

\bibitem[{\citenamefont{Khomskii}(2009)}]{Khomskii2009}
\bibinfo{author}{\bibfnamefont{D.~I.} \bibnamefont{Khomskii}},
  \bibinfo{journal}{Physics} \textbf{\bibinfo{volume}{2}}
  (\bibinfo{year}{2009}).

\bibitem[{\citenamefont{Tokura et~al.}(2014)\citenamefont{Tokura, Seki, and
  Nagaosa}}]{Tokura2014}
\bibinfo{author}{\bibfnamefont{Y.}~\bibnamefont{Tokura}},
  \bibinfo{author}{\bibfnamefont{S.}~\bibnamefont{Seki}}, \bibnamefont{and}
  \bibinfo{author}{\bibfnamefont{N.}~\bibnamefont{Nagaosa}},
  \bibinfo{journal}{Reports on Progress in Physics}
  \textbf{\bibinfo{volume}{77}}, \bibinfo{pages}{076501}
  (\bibinfo{year}{2014}).

\bibitem[{\citenamefont{Jia et~al.}(2006)\citenamefont{Jia, Onoda, Nagaosa, and
  Han}}]{Jia2006}
\bibinfo{author}{\bibfnamefont{C.}~\bibnamefont{Jia}},
  \bibinfo{author}{\bibfnamefont{S.}~\bibnamefont{Onoda}},
  \bibinfo{author}{\bibfnamefont{N.}~\bibnamefont{Nagaosa}}, \bibnamefont{and}
  \bibinfo{author}{\bibfnamefont{J.~H.} \bibnamefont{Han}},
  \bibinfo{journal}{Phys. Rev. B} \textbf{\bibinfo{volume}{74}},
  \bibinfo{pages}{224444} (\bibinfo{year}{2006}).

\bibitem[{\citenamefont{Jia et~al.}(2007)\citenamefont{Jia, Onoda, Nagaosa, and
  Han}}]{Jia2007}
\bibinfo{author}{\bibfnamefont{C.}~\bibnamefont{Jia}},
  \bibinfo{author}{\bibfnamefont{S.}~\bibnamefont{Onoda}},
  \bibinfo{author}{\bibfnamefont{N.}~\bibnamefont{Nagaosa}}, \bibnamefont{and}
  \bibinfo{author}{\bibfnamefont{J.~H.} \bibnamefont{Han}},
  \bibinfo{journal}{Phys. Rev. B} \textbf{\bibinfo{volume}{76}},
  \bibinfo{pages}{144424} (\bibinfo{year}{2007}).

\bibitem[{\citenamefont{hisa Arima}(2007)}]{Arima2007}
\bibinfo{author}{\bibfnamefont{T.}~\bibnamefont{hisa Arima}},
  \bibinfo{journal}{Journal of the Physical Society of Japan}
  \textbf{\bibinfo{volume}{76}}, \bibinfo{pages}{073702}
  (\bibinfo{year}{2007}).

\bibitem[{\citenamefont{Bos et~al.}(2008)\citenamefont{Bos, Colin, and
  Palstra}}]{Bos2008}
\bibinfo{author}{\bibfnamefont{J.-W.~G.} \bibnamefont{Bos}},
  \bibinfo{author}{\bibfnamefont{C.~V.} \bibnamefont{Colin}}, \bibnamefont{and}
  \bibinfo{author}{\bibfnamefont{T.~T.~M.} \bibnamefont{Palstra}},
  \bibinfo{journal}{Phys. Rev. B} \textbf{\bibinfo{volume}{78}},
  \bibinfo{pages}{094416} (\bibinfo{year}{2008}).

\bibitem[{\citenamefont{Belesi et~al.}(2012)\citenamefont{Belesi,
  Rousochatzakis, Abid, R\"o\ss{}ler, Berger, and Ansermet}}]{Belesi2012}
\bibinfo{author}{\bibfnamefont{M.}~\bibnamefont{Belesi}},
  \bibinfo{author}{\bibfnamefont{I.}~\bibnamefont{Rousochatzakis}},
  \bibinfo{author}{\bibfnamefont{M.}~\bibnamefont{Abid}},
  \bibinfo{author}{\bibfnamefont{U.~K.} \bibnamefont{R\"o\ss{}ler}},
  \bibinfo{author}{\bibfnamefont{H.}~\bibnamefont{Berger}}, \bibnamefont{and}
  \bibinfo{author}{\bibfnamefont{J.-P.} \bibnamefont{Ansermet}},
  \bibinfo{journal}{Phys. Rev. B} \textbf{\bibinfo{volume}{85}},
  \bibinfo{pages}{224413} (\bibinfo{year}{2012}).

\bibitem[{\citenamefont{Mochizuki and Seki}(2013)}]{Mochizuki2013}
\bibinfo{author}{\bibfnamefont{M.}~\bibnamefont{Mochizuki}} \bibnamefont{and}
  \bibinfo{author}{\bibfnamefont{S.}~\bibnamefont{Seki}},
  \bibinfo{journal}{Phys. Rev. B} \textbf{\bibinfo{volume}{87}},
  \bibinfo{pages}{134403} (\bibinfo{year}{2013}).

\bibitem[{\citenamefont{Mochizuki}(2012)}]{Mochizuki2012}
\bibinfo{author}{\bibfnamefont{M.}~\bibnamefont{Mochizuki}},
  \bibinfo{journal}{Phys. Rev. Lett.} \textbf{\bibinfo{volume}{108}},
  \bibinfo{pages}{017601} (\bibinfo{year}{2012}).

\bibitem[{\citenamefont{Maisuradze et~al.}(2012)\citenamefont{Maisuradze,
  Shengelaya, Berger, Djoki\ifmmode~\acute{c}\else \'{c}\fi{}, and
  Keller}}]{Maisurandze2012}
\bibinfo{author}{\bibfnamefont{A.}~\bibnamefont{Maisuradze}},
  \bibinfo{author}{\bibfnamefont{A.}~\bibnamefont{Shengelaya}},
  \bibinfo{author}{\bibfnamefont{H.}~\bibnamefont{Berger}},
  \bibinfo{author}{\bibfnamefont{D.~M.}
  \bibnamefont{Djoki\ifmmode~\acute{c}\else \'{c}\fi{}}}, \bibnamefont{and}
  \bibinfo{author}{\bibfnamefont{H.}~\bibnamefont{Keller}},
  \bibinfo{journal}{Phys. Rev. Lett.} \textbf{\bibinfo{volume}{108}},
  \bibinfo{pages}{247211} (\bibinfo{year}{2012}).

\bibitem[{\citenamefont{Maisuradze et~al.}(2011)\citenamefont{Maisuradze,
  Guguchia, Graneli, R\o{}nnow, Berger, and Keller}}]{Maisurandze2011}
\bibinfo{author}{\bibfnamefont{A.}~\bibnamefont{Maisuradze}},
  \bibinfo{author}{\bibfnamefont{Z.}~\bibnamefont{Guguchia}},
  \bibinfo{author}{\bibfnamefont{B.}~\bibnamefont{Graneli}},
  \bibinfo{author}{\bibfnamefont{H.~M.} \bibnamefont{R\o{}nnow}},
  \bibinfo{author}{\bibfnamefont{H.}~\bibnamefont{Berger}}, \bibnamefont{and}
  \bibinfo{author}{\bibfnamefont{H.}~\bibnamefont{Keller}},
  \bibinfo{journal}{Phys. Rev. B} \textbf{\bibinfo{volume}{84}},
  \bibinfo{pages}{064433} (\bibinfo{year}{2011}).

\bibitem[{\citenamefont{Omrani et~al.}(2014)\citenamefont{Omrani, White,
  Pr\ifmmode~\check{s}\else \v{s}\fi{}a, \ifmmode \check{Z}\else
  \v{Z}\fi{}ivkovi\ifmmode~\acute{c}\else \'{c}\fi{}, Berger, Magrez, Liu, Han,
  and R\o{}nnow}}]{Omrani2014}
\bibinfo{author}{\bibfnamefont{A.~A.} \bibnamefont{Omrani}},
  \bibinfo{author}{\bibfnamefont{J.~S.} \bibnamefont{White}},
  \bibinfo{author}{\bibfnamefont{K.}~\bibnamefont{Pr\ifmmode~\check{s}\else
  \v{s}\fi{}a}}, \bibinfo{author}{\bibfnamefont{I.}~\bibnamefont{\ifmmode
  \check{Z}\else \v{Z}\fi{}ivkovi\ifmmode~\acute{c}\else \'{c}\fi{}}},
  \bibinfo{author}{\bibfnamefont{H.}~\bibnamefont{Berger}},
  \bibinfo{author}{\bibfnamefont{A.}~\bibnamefont{Magrez}},
  \bibinfo{author}{\bibfnamefont{Y.-H.} \bibnamefont{Liu}},
  \bibinfo{author}{\bibfnamefont{J.~H.} \bibnamefont{Han}}, \bibnamefont{and}
  \bibinfo{author}{\bibfnamefont{H.~M.} \bibnamefont{R\o{}nnow}},
  \bibinfo{journal}{Phys. Rev. B} \textbf{\bibinfo{volume}{89}},
  \bibinfo{pages}{064406} (\bibinfo{year}{2014}).

\bibitem[{\citenamefont{Ruff et~al.}(2015)\citenamefont{Ruff, Lunkenheimer,
  Loidl, Berger, and Krohns}}]{Ruff2015}
\bibinfo{author}{\bibfnamefont{E.}~\bibnamefont{Ruff}},
  \bibinfo{author}{\bibfnamefont{P.}~\bibnamefont{Lunkenheimer}},
  \bibinfo{author}{\bibfnamefont{A.}~\bibnamefont{Loidl}},
  \bibinfo{author}{\bibfnamefont{H.}~\bibnamefont{Berger}}, \bibnamefont{and}
  \bibinfo{author}{\bibfnamefont{S.}~\bibnamefont{Krohns}},
  \bibinfo{journal}{Scientific Reports} \textbf{\bibinfo{volume}{5}}
  (\bibinfo{year}{2015}).

\bibitem[{\citenamefont{Janson et~al.}(2014)\citenamefont{Janson,
  Rousochatzakis, Tsirlin, Belesi, Leonov, Robler, van~den Brink, and
  Rosner}}]{Janson2014}
\bibinfo{author}{\bibfnamefont{O.}~\bibnamefont{Janson}},
  \bibinfo{author}{\bibfnamefont{I.}~\bibnamefont{Rousochatzakis}},
  \bibinfo{author}{\bibfnamefont{A.~A.} \bibnamefont{Tsirlin}},
  \bibinfo{author}{\bibfnamefont{M.}~\bibnamefont{Belesi}},
  \bibinfo{author}{\bibfnamefont{A.~A.} \bibnamefont{Leonov}},
  \bibinfo{author}{\bibfnamefont{U.~K.} \bibnamefont{Robler}},
  \bibinfo{author}{\bibfnamefont{J.}~\bibnamefont{van~den Brink}},
  \bibnamefont{and} \bibinfo{author}{\bibfnamefont{H.}~\bibnamefont{Rosner}},
  \bibinfo{journal}{Nature Comm.}  (\bibinfo{year}{2014}).

\bibitem[{\citenamefont{Miller et~al.}(2010)\citenamefont{Miller, Xu., Berger,
  Knowles, Arenas, Meisel, and Tanner}}]{Miller2010}
\bibinfo{author}{\bibfnamefont{K.~H.} \bibnamefont{Miller}},
  \bibinfo{author}{\bibfnamefont{X.~S.} \bibnamefont{Xu.}},
  \bibinfo{author}{\bibfnamefont{H.}~\bibnamefont{Berger}},
  \bibinfo{author}{\bibfnamefont{E.~S.} \bibnamefont{Knowles}},
  \bibinfo{author}{\bibfnamefont{D.~J.} \bibnamefont{Arenas}},
  \bibinfo{author}{\bibfnamefont{M.~W.} \bibnamefont{Meisel}},
  \bibnamefont{and} \bibinfo{author}{\bibfnamefont{D.~B.}
  \bibnamefont{Tanner}}, \bibinfo{journal}{Phys. Rev. B}
  \textbf{\bibinfo{volume}{82}}, \bibinfo{pages}{144107}
  (\bibinfo{year}{2010}).

\bibitem[{\citenamefont{Versteeg et~al.}(2016)\citenamefont{Versteeg, Vergara,
  Sch\"afer, Bischoff, Aqeel, Palstra, Gr\"uninger, and van
  Loosdrecht}}]{Versteeg2016}
\bibinfo{author}{\bibfnamefont{R.~B.} \bibnamefont{Versteeg}},
  \bibinfo{author}{\bibfnamefont{I.}~\bibnamefont{Vergara}},
  \bibinfo{author}{\bibfnamefont{S.~D.} \bibnamefont{Sch\"afer}},
  \bibinfo{author}{\bibfnamefont{D.}~\bibnamefont{Bischoff}},
  \bibinfo{author}{\bibfnamefont{A.}~\bibnamefont{Aqeel}},
  \bibinfo{author}{\bibfnamefont{T.~T.~M.} \bibnamefont{Palstra}},
  \bibinfo{author}{\bibfnamefont{M.}~\bibnamefont{Gr\"uninger}},
  \bibnamefont{and} \bibinfo{author}{\bibfnamefont{P.~H.~M.} \bibnamefont{van
  Loosdrecht}}, \bibinfo{journal}{Phys. Rev. B} \textbf{\bibinfo{volume}{94}},
  \bibinfo{pages}{094409} (\bibinfo{year}{2016}).

\bibitem[{\citenamefont{Szaller et~al.}(2013)\citenamefont{Szaller, Bord\'acs,
  and K\'ezsm\'arki}}]{Szaller2013}
\bibinfo{author}{\bibfnamefont{D.}~\bibnamefont{Szaller}},
  \bibinfo{author}{\bibfnamefont{S.}~\bibnamefont{Bord\'acs}},
  \bibnamefont{and}
  \bibinfo{author}{\bibfnamefont{I.}~\bibnamefont{K\'ezsm\'arki}},
  \bibinfo{journal}{Phys. Rev. B} \textbf{\bibinfo{volume}{87}},
  \bibinfo{pages}{014421} (\bibinfo{year}{2013}).

\bibitem[{\citenamefont{Okamura et~al.}(2015)\citenamefont{Okamura, Kagawa,
  Seki, Kubota, Kawasaki, and Tokura}}]{Okamura2015}
\bibinfo{author}{\bibfnamefont{Y.}~\bibnamefont{Okamura}},
  \bibinfo{author}{\bibfnamefont{F.}~\bibnamefont{Kagawa}},
  \bibinfo{author}{\bibfnamefont{S.}~\bibnamefont{Seki}},
  \bibinfo{author}{\bibfnamefont{M.}~\bibnamefont{Kubota}},
  \bibinfo{author}{\bibfnamefont{M.}~\bibnamefont{Kawasaki}}, \bibnamefont{and}
  \bibinfo{author}{\bibfnamefont{Y.}~\bibnamefont{Tokura}},
  \bibinfo{journal}{Phys. Rev. Lett.} \textbf{\bibinfo{volume}{114}},
  \bibinfo{pages}{197202} (\bibinfo{year}{2015}).

\bibitem[{\citenamefont{Mochizuki}(2015)}]{Mochizuki2015}
\bibinfo{author}{\bibfnamefont{M.}~\bibnamefont{Mochizuki}},
  \bibinfo{journal}{Phys. Rev. Lett.} \textbf{\bibinfo{volume}{114}},
  \bibinfo{pages}{197203} (\bibinfo{year}{2015}).

\bibitem[{\citenamefont{Schwarze et~al.}(2015)\citenamefont{Schwarze, Waizner,
  Garst, Bauer, Stasinopoulos, Berger, Pfleiderer, and
  Grundler}}]{Schwarze2015}
\bibinfo{author}{\bibfnamefont{T.}~\bibnamefont{Schwarze}},
  \bibinfo{author}{\bibfnamefont{J.}~\bibnamefont{Waizner}},
  \bibinfo{author}{\bibfnamefont{M.}~\bibnamefont{Garst}},
  \bibinfo{author}{\bibfnamefont{A.}~\bibnamefont{Bauer}},
  \bibinfo{author}{\bibfnamefont{I.}~\bibnamefont{Stasinopoulos}},
  \bibinfo{author}{\bibfnamefont{H.}~\bibnamefont{Berger}},
  \bibinfo{author}{\bibfnamefont{C.}~\bibnamefont{Pfleiderer}},
  \bibnamefont{and} \bibinfo{author}{\bibfnamefont{D.}~\bibnamefont{Grundler}},
  \bibinfo{journal}{Nature Mater.} \textbf{\bibinfo{volume}{14}},
  \bibinfo{pages}{478} (\bibinfo{year}{2015}).

\bibitem[{\citenamefont{Gnezdilov et~al.}(2010)\citenamefont{Gnezdilov,
  Lamonova, Pashkevich, Lemmens, Berger, Bussy, and
  Gnatchenko}}]{Gnezdilov2010}
\bibinfo{author}{\bibfnamefont{V.~P.} \bibnamefont{Gnezdilov}},
  \bibinfo{author}{\bibfnamefont{Y.}~\bibnamefont{Lamonova}},
  \bibinfo{author}{\bibfnamefont{G.}~\bibnamefont{Pashkevich}},
  \bibinfo{author}{\bibfnamefont{P.}~\bibnamefont{Lemmens}},
  \bibinfo{author}{\bibfnamefont{H.}~\bibnamefont{Berger}},
  \bibinfo{author}{\bibfnamefont{F.}~\bibnamefont{Bussy}}, \bibnamefont{and}
  \bibinfo{author}{\bibfnamefont{S.~L.} \bibnamefont{Gnatchenko}},
  \bibinfo{journal}{Low Temperature Physics} \textbf{\bibinfo{volume}{36}}
  (\bibinfo{year}{2010}).

\bibitem[{\citenamefont{Ozerov et~al.}(2014)\citenamefont{Ozerov, Romh\'anyi,
  Belesi, Berger, Ansermet, van~den Brink, Wosnitza, Zvyagin, and
  Rousochatzakis}}]{Ozerov2014}
\bibinfo{author}{\bibfnamefont{M.}~\bibnamefont{Ozerov}},
  \bibinfo{author}{\bibfnamefont{J.}~\bibnamefont{Romh\'anyi}},
  \bibinfo{author}{\bibfnamefont{M.}~\bibnamefont{Belesi}},
  \bibinfo{author}{\bibfnamefont{H.}~\bibnamefont{Berger}},
  \bibinfo{author}{\bibfnamefont{J.-P.} \bibnamefont{Ansermet}},
  \bibinfo{author}{\bibfnamefont{J.}~\bibnamefont{van~den Brink}},
  \bibinfo{author}{\bibfnamefont{J.}~\bibnamefont{Wosnitza}},
  \bibinfo{author}{\bibfnamefont{S.~A.} \bibnamefont{Zvyagin}},
  \bibnamefont{and}
  \bibinfo{author}{\bibfnamefont{I.}~\bibnamefont{Rousochatzakis}},
  \bibinfo{journal}{Phys. Rev. Lett.} \textbf{\bibinfo{volume}{113}},
  \bibinfo{pages}{157205} (\bibinfo{year}{2014}).

\bibitem[{\citenamefont{Momma and Izumi}(2011)}]{Momma2011}
\bibinfo{author}{\bibfnamefont{K.}~\bibnamefont{Momma}} \bibnamefont{and}
  \bibinfo{author}{\bibfnamefont{F.}~\bibnamefont{Izumi}},
  \bibinfo{journal}{Journal of Applied Crystallography}
  \textbf{\bibinfo{volume}{44}}, \bibinfo{pages}{1272} (\bibinfo{year}{2011}).

\bibitem[{\citenamefont{Effenberger and Pertlik}(1986)}]{Effenberger1986}
\bibinfo{author}{\bibfnamefont{H.}~\bibnamefont{Effenberger}} \bibnamefont{and}
  \bibinfo{author}{\bibfnamefont{F.}~\bibnamefont{Pertlik}},
  \bibinfo{journal}{Chemical Monthly} \textbf{\bibinfo{volume}{117}},
  \bibinfo{pages}{887} (\bibinfo{year}{1986}).

\bibitem[{\citenamefont{Belesi et~al.}(2010)\citenamefont{Belesi,
  Rousochatzakis, Wu, Berger, Shvets, Mila, and Ansermet}}]{Belesi2010}
\bibinfo{author}{\bibfnamefont{M.}~\bibnamefont{Belesi}},
  \bibinfo{author}{\bibfnamefont{I.}~\bibnamefont{Rousochatzakis}},
  \bibinfo{author}{\bibfnamefont{H.~C.} \bibnamefont{Wu}},
  \bibinfo{author}{\bibfnamefont{H.}~\bibnamefont{Berger}},
  \bibinfo{author}{\bibfnamefont{I.~V.} \bibnamefont{Shvets}},
  \bibinfo{author}{\bibfnamefont{F.}~\bibnamefont{Mila}}, \bibnamefont{and}
  \bibinfo{author}{\bibfnamefont{J.~P.} \bibnamefont{Ansermet}},
  \bibinfo{journal}{Phys. Rev. B} \textbf{\bibinfo{volume}{82}},
  \bibinfo{pages}{094422} (\bibinfo{year}{2010}).

\bibitem[{\citenamefont{Dzyaloshinsky}(1958)}]{Dz1958}
\bibinfo{author}{\bibfnamefont{I.}~\bibnamefont{Dzyaloshinsky}},
  \bibinfo{journal}{Journal of Physics and Chemistry of Solids}
  \textbf{\bibinfo{volume}{4}}, \bibinfo{pages}{241 } (\bibinfo{year}{1958}).

\bibitem[{\citenamefont{Moriya}(1960)}]{Moriya1960}
\bibinfo{author}{\bibfnamefont{T.}~\bibnamefont{Moriya}},
  \bibinfo{journal}{Phys. Rev.} \textbf{\bibinfo{volume}{120}},
  \bibinfo{pages}{91} (\bibinfo{year}{1960}).

\bibitem[{\citenamefont{Portnichenko et~al.}(2016)\citenamefont{Portnichenko,
  Romhayni, Onykiienko, Henschel, Schmidt, Cameron, Surmach, Lim, Park,
  Schneidewind et~al.}}]{Portnichenko2016}
\bibinfo{author}{\bibfnamefont{P.~Y.} \bibnamefont{Portnichenko}},
  \bibinfo{author}{\bibfnamefont{J.}~\bibnamefont{Romhayni}},
  \bibinfo{author}{\bibfnamefont{Y.~A.} \bibnamefont{Onykiienko}},
  \bibinfo{author}{\bibfnamefont{A.}~\bibnamefont{Henschel}},
  \bibinfo{author}{\bibfnamefont{M.}~\bibnamefont{Schmidt}},
  \bibinfo{author}{\bibfnamefont{A.~S.} \bibnamefont{Cameron}},
  \bibinfo{author}{\bibfnamefont{M.~A.} \bibnamefont{Surmach}},
  \bibinfo{author}{\bibfnamefont{J.~A.} \bibnamefont{Lim}},
  \bibinfo{author}{\bibfnamefont{J.~T.} \bibnamefont{Park}},
  \bibinfo{author}{\bibfnamefont{A.}~\bibnamefont{Schneidewind}},
  \bibnamefont{et~al.}, \bibinfo{journal}{Nature Commun.}
  \textbf{\bibinfo{volume}{7}}, \bibinfo{pages}{10725} (\bibinfo{year}{2016}).

\bibitem[{\citenamefont{Tucker et~al.}(2016)\citenamefont{Tucker, White,
  Romh\'anyi, Szaller, K\'ezsm\'arki, Roessli, Stuhr, Magrez, Groitl, Babkevich
  et~al.}}]{Tucker2016}
\bibinfo{author}{\bibfnamefont{G.~S.} \bibnamefont{Tucker}},
  \bibinfo{author}{\bibfnamefont{J.~S.} \bibnamefont{White}},
  \bibinfo{author}{\bibfnamefont{J.}~\bibnamefont{Romh\'anyi}},
  \bibinfo{author}{\bibfnamefont{D.}~\bibnamefont{Szaller}},
  \bibinfo{author}{\bibfnamefont{I.}~\bibnamefont{K\'ezsm\'arki}},
  \bibinfo{author}{\bibfnamefont{B.}~\bibnamefont{Roessli}},
  \bibinfo{author}{\bibfnamefont{U.}~\bibnamefont{Stuhr}},
  \bibinfo{author}{\bibfnamefont{A.}~\bibnamefont{Magrez}},
  \bibinfo{author}{\bibfnamefont{F.}~\bibnamefont{Groitl}},
  \bibinfo{author}{\bibfnamefont{P.}~\bibnamefont{Babkevich}},
  \bibnamefont{et~al.}, \bibinfo{journal}{Phys. Rev. B}
  \textbf{\bibinfo{volume}{93}}, \bibinfo{pages}{054401}
  (\bibinfo{year}{2016}).

\bibitem[{\citenamefont{Romh\'anyi et~al.}(2014)\citenamefont{Romh\'anyi,
  van~den Brink, and Rousochatzakis}}]{Romhanyi2014}
\bibinfo{author}{\bibfnamefont{J.}~\bibnamefont{Romh\'anyi}},
  \bibinfo{author}{\bibfnamefont{J.}~\bibnamefont{van~den Brink}},
  \bibnamefont{and}
  \bibinfo{author}{\bibfnamefont{I.}~\bibnamefont{Rousochatzakis}},
  \bibinfo{journal}{Phys. Rev. B} \textbf{\bibinfo{volume}{90}},
  \bibinfo{pages}{140404} (\bibinfo{year}{2014}).

\bibitem[{\citenamefont{Chizhikov and Dmitrienko}(2015)}]{Chizikov2015}
\bibinfo{author}{\bibfnamefont{V.}~\bibnamefont{Chizhikov}} \bibnamefont{and}
  \bibinfo{author}{\bibfnamefont{V.}~\bibnamefont{Dmitrienko}},
  \bibinfo{journal}{Journal of Magnetism and Magnetic Materials}
  \textbf{\bibinfo{volume}{382}}, \bibinfo{pages}{142 } (\bibinfo{year}{2015}).

\bibitem[{\citenamefont{Adams et~al.}(2012)\citenamefont{Adams, Chacon, Wagner,
  Bauer, Brandl, Pedersen, Berger, Lemmens, and Pfleiderer}}]{Adams2012}
\bibinfo{author}{\bibfnamefont{T.}~\bibnamefont{Adams}},
  \bibinfo{author}{\bibfnamefont{A.}~\bibnamefont{Chacon}},
  \bibinfo{author}{\bibfnamefont{M.}~\bibnamefont{Wagner}},
  \bibinfo{author}{\bibfnamefont{A.}~\bibnamefont{Bauer}},
  \bibinfo{author}{\bibfnamefont{G.}~\bibnamefont{Brandl}},
  \bibinfo{author}{\bibfnamefont{B.}~\bibnamefont{Pedersen}},
  \bibinfo{author}{\bibfnamefont{H.}~\bibnamefont{Berger}},
  \bibinfo{author}{\bibfnamefont{P.}~\bibnamefont{Lemmens}}, \bibnamefont{and}
  \bibinfo{author}{\bibfnamefont{C.}~\bibnamefont{Pfleiderer}},
  \bibinfo{journal}{Phys. Rev. Lett.} \textbf{\bibinfo{volume}{108}},
  \bibinfo{pages}{237204} (\bibinfo{year}{2012}).

\bibitem[{\citenamefont{Levati\ifmmode~\acute{c}\else \'{c}\fi{}
  et~al.}(2014)\citenamefont{Levati\ifmmode~\acute{c}\else \'{c}\fi{},
  \ifmmode~\check{S}\else \v{S}\fi{}urija, Berger, and \ifmmode \check{Z}\else
  \v{Z}\fi{}ivkovi\ifmmode~\acute{c}\else \'{c}\fi{}}}]{Levatic2014}
\bibinfo{author}{\bibfnamefont{I.}~\bibnamefont{Levati\ifmmode~\acute{c}\else
  \'{c}\fi{}}},
  \bibinfo{author}{\bibfnamefont{V.}~\bibnamefont{\ifmmode~\check{S}\else
  \v{S}\fi{}urija}}, \bibinfo{author}{\bibfnamefont{H.}~\bibnamefont{Berger}},
  \bibnamefont{and} \bibinfo{author}{\bibfnamefont{I.}~\bibnamefont{\ifmmode
  \check{Z}\else \v{Z}\fi{}ivkovi\ifmmode~\acute{c}\else \'{c}\fi{}}},
  \bibinfo{journal}{Phys. Rev. B} \textbf{\bibinfo{volume}{90}},
  \bibinfo{pages}{224412} (\bibinfo{year}{2014}).

\bibitem[{\citenamefont{Panella et~al.}(2017)\citenamefont{Panella, Trump,
  Marcus, and McQueen}}]{Panella2017}
\bibinfo{author}{\bibfnamefont{J.}~\bibnamefont{Panella}},
  \bibinfo{author}{\bibfnamefont{B.~A.} \bibnamefont{Trump}},
  \bibinfo{author}{\bibfnamefont{G.~G.} \bibnamefont{Marcus}},
  \bibnamefont{and} \bibinfo{author}{\bibfnamefont{T.~M.}
  \bibnamefont{McQueen}}, \bibinfo{journal}{In Preparation}
  (\bibinfo{year}{2017}).

\bibitem[{\citenamefont{Laurita et~al.}(2016)\citenamefont{Laurita, Cheng,
  Barkhouser, Neumann, and Armitage}}]{Laurita2016}
\bibinfo{author}{\bibfnamefont{N.~J.} \bibnamefont{Laurita}},
  \bibinfo{author}{\bibfnamefont{B.}~\bibnamefont{Cheng}},
  \bibinfo{author}{\bibfnamefont{R.}~\bibnamefont{Barkhouser}},
  \bibinfo{author}{\bibfnamefont{V.~A.} \bibnamefont{Neumann}},
  \bibnamefont{and} \bibinfo{author}{\bibfnamefont{N.~P.}
  \bibnamefont{Armitage}}, \bibinfo{journal}{Journal of Infrared, Millimeter,
  and Terahertz Waves} pp. \bibinfo{pages}{1--9} (\bibinfo{year}{2016}).

\bibitem[{\citenamefont{Press et~al.}(2007)\citenamefont{Press, Teukolsky,
  Vetterling, and Flannery}}]{Press2007}
\bibinfo{author}{\bibfnamefont{W.~H.} \bibnamefont{Press}},
  \bibinfo{author}{\bibfnamefont{S.~A.} \bibnamefont{Teukolsky}},
  \bibinfo{author}{\bibfnamefont{W.~T.} \bibnamefont{Vetterling}},
  \bibnamefont{and} \bibinfo{author}{\bibfnamefont{B.~P.}
  \bibnamefont{Flannery}}, \emph{\bibinfo{title}{Numerical Recipes 3rd Edition:
  The Art of Scientific Computing}} (\bibinfo{publisher}{Cambridge University
  Press}, \bibinfo{address}{New York, NY, USA}, \bibinfo{year}{2007}),
  \bibinfo{edition}{3rd} ed.

\bibitem[{\citenamefont{Jones}(1941)}]{Jones1941}
\bibinfo{author}{\bibfnamefont{R.~C.} \bibnamefont{Jones}},
  \bibinfo{journal}{J. Opt. Soc. Am.} \textbf{\bibinfo{volume}{31}},
  \bibinfo{pages}{488} (\bibinfo{year}{1941}).

\bibitem[{\citenamefont{Armitage}(2014)}]{Armitage2014}
\bibinfo{author}{\bibfnamefont{N.~P.} \bibnamefont{Armitage}},
  \bibinfo{journal}{Phys. Rev. B} \textbf{\bibinfo{volume}{90}},
  \bibinfo{pages}{035135} (\bibinfo{year}{2014}).

\bibitem[{\citenamefont{Morris et~al.}(2012)\citenamefont{Morris, Aguilar,
  Stier, and Armitage}}]{Morris2012}
\bibinfo{author}{\bibfnamefont{C.~M.} \bibnamefont{Morris}},
  \bibinfo{author}{\bibfnamefont{R.~V.} \bibnamefont{Aguilar}},
  \bibinfo{author}{\bibfnamefont{A.~V.} \bibnamefont{Stier}}, \bibnamefont{and}
  \bibinfo{author}{\bibfnamefont{N.~P.} \bibnamefont{Armitage}},
  \bibinfo{journal}{Opt. Express} \textbf{\bibinfo{volume}{20}},
  \bibinfo{pages}{12303} (\bibinfo{year}{2012}).

\bibitem[{\citenamefont{Arnold et~al.}(2014)\citenamefont{Arnold, Bilheux,
  Borreguero, Buts, Campbell, Chapon, Doucet, Draper, Leal, Gigg
  et~al.}}]{Arnold2014}
\bibinfo{author}{\bibfnamefont{O.}~\bibnamefont{Arnold}},
  \bibinfo{author}{\bibfnamefont{J.}~\bibnamefont{Bilheux}},
  \bibinfo{author}{\bibfnamefont{J.}~\bibnamefont{Borreguero}},
  \bibinfo{author}{\bibfnamefont{A.}~\bibnamefont{Buts}},
  \bibinfo{author}{\bibfnamefont{S.}~\bibnamefont{Campbell}},
  \bibinfo{author}{\bibfnamefont{L.}~\bibnamefont{Chapon}},
  \bibinfo{author}{\bibfnamefont{M.}~\bibnamefont{Doucet}},
  \bibinfo{author}{\bibfnamefont{N.}~\bibnamefont{Draper}},
  \bibinfo{author}{\bibfnamefont{R.~F.} \bibnamefont{Leal}},
  \bibinfo{author}{\bibfnamefont{M.}~\bibnamefont{Gigg}}, \bibnamefont{et~al.},
  \bibinfo{journal}{Nuclear Instruments and Methods in Physics Research Section
  A: Accelerators, Spectrometers, Detectors and Associated Equipment}
  \textbf{\bibinfo{volume}{764}}, \bibinfo{pages}{156 } (\bibinfo{year}{2014}).

\bibitem[{\citenamefont{Ewings et~al.}(2016)\citenamefont{Ewings, Buts, Le, van
  Duijn, Bustinduy, and Perring}}]{Ewings2016}
\bibinfo{author}{\bibfnamefont{R.}~\bibnamefont{Ewings}},
  \bibinfo{author}{\bibfnamefont{A.}~\bibnamefont{Buts}},
  \bibinfo{author}{\bibfnamefont{M.}~\bibnamefont{Le}},
  \bibinfo{author}{\bibfnamefont{J.}~\bibnamefont{van Duijn}},
  \bibinfo{author}{\bibfnamefont{I.}~\bibnamefont{Bustinduy}},
  \bibnamefont{and} \bibinfo{author}{\bibfnamefont{T.}~\bibnamefont{Perring}},
  \bibinfo{journal}{Nuclear Instruments and Methods in Physics Research Section
  A: Accelerators, Spectrometers, Detectors and Associated Equipment}
  \textbf{\bibinfo{volume}{834}}, \bibinfo{pages}{132 } (\bibinfo{year}{2016}).

\bibitem[{\citenamefont{Marcus~\textit{et al.}}(2017)}]{Marcus2017}
\bibinfo{author}{\bibfnamefont{G.~G.} \bibnamefont{Marcus~\textit{et al.}}},
  \bibinfo{journal}{In Preparation.}  (\bibinfo{year}{2017}).

\bibitem[{\citenamefont{Freiser}(1968)}]{Freiser1968}
\bibinfo{author}{\bibfnamefont{M.}~\bibnamefont{Freiser}},
  \bibinfo{journal}{IEEE Transactions on Magnetics}
  \textbf{\bibinfo{volume}{4}}, \bibinfo{pages}{152} (\bibinfo{year}{1968}).

\bibitem[{\citenamefont{Woodford et~al.}(2007)\citenamefont{Woodford, Bringer,
  and Blügel}}]{Woodford2007}
\bibinfo{author}{\bibfnamefont{S.~R.} \bibnamefont{Woodford}},
  \bibinfo{author}{\bibfnamefont{A.}~\bibnamefont{Bringer}}, \bibnamefont{and}
  \bibinfo{author}{\bibfnamefont{S.}~\bibnamefont{Blügel}},
  \bibinfo{journal}{Journal of Applied Physics} \textbf{\bibinfo{volume}{101}}
  (\bibinfo{year}{2007}).

\bibitem[{\citenamefont{Glazer and Cox}(2006)}]{Glazer2006}
\bibinfo{author}{\bibfnamefont{A.}~\bibnamefont{Glazer}} \bibnamefont{and}
  \bibinfo{author}{\bibfnamefont{K.~G.} \bibnamefont{Cox}},
  \emph{\bibinfo{title}{Clasical linear crystal optics in International Tables
  for Crystallography}}, vol.~\bibinfo{volume}{D} (\bibinfo{publisher}{John
  Wiley \& Sons, Ltd.}, \bibinfo{year}{2006}).

\bibitem[{\citenamefont{Barron}(2009)}]{Barron2009}
\bibinfo{author}{\bibfnamefont{L.~D.} \bibnamefont{Barron}},
  \emph{\bibinfo{title}{Molecular Light Scattering and Optical Activity}}
  (\bibinfo{publisher}{Cambridge University Press, New York},
  \bibinfo{year}{2009}).

\bibitem[{\citenamefont{\ifmmode \check{Z}\else
  \v{Z}\fi{}ivkovi\ifmmode~\acute{c}\else \'{c}\fi{}
  et~al.}(2012)\citenamefont{\ifmmode \check{Z}\else
  \v{Z}\fi{}ivkovi\ifmmode~\acute{c}\else \'{c}\fi{},
  Paji\ifmmode~\acute{c}\else \'{c}\fi{}, Ivek, and Berger}}]{Zivkovic2012}
\bibinfo{author}{\bibfnamefont{I.}~\bibnamefont{\ifmmode \check{Z}\else
  \v{Z}\fi{}ivkovi\ifmmode~\acute{c}\else \'{c}\fi{}}},
  \bibinfo{author}{\bibfnamefont{D.}~\bibnamefont{Paji\ifmmode~\acute{c}\else
  \'{c}\fi{}}}, \bibinfo{author}{\bibfnamefont{T.}~\bibnamefont{Ivek}},
  \bibnamefont{and} \bibinfo{author}{\bibfnamefont{H.}~\bibnamefont{Berger}},
  \bibinfo{journal}{Phys. Rev. B} \textbf{\bibinfo{volume}{85}},
  \bibinfo{pages}{224402} (\bibinfo{year}{2012}).

\bibitem[{\citenamefont{\ifmmode \check{Z}\else
  \v{Z}\fi{}ivkovi\ifmmode~\acute{c}\else \'{c}\fi{}
  et~al.}(2014)\citenamefont{\ifmmode \check{Z}\else
  \v{Z}\fi{}ivkovi\ifmmode~\acute{c}\else \'{c}\fi{}, White, R\o{}nnow,
  Pr\ifmmode~\check{s}\else \v{s}\fi{}a, and Berger}}]{Zivkovic2014}
\bibinfo{author}{\bibfnamefont{I.}~\bibnamefont{\ifmmode \check{Z}\else
  \v{Z}\fi{}ivkovi\ifmmode~\acute{c}\else \'{c}\fi{}}},
  \bibinfo{author}{\bibfnamefont{J.~S.} \bibnamefont{White}},
  \bibinfo{author}{\bibfnamefont{H.~M.} \bibnamefont{R\o{}nnow}},
  \bibinfo{author}{\bibfnamefont{K.}~\bibnamefont{Pr\ifmmode~\check{s}\else
  \v{s}\fi{}a}}, \bibnamefont{and}
  \bibinfo{author}{\bibfnamefont{H.}~\bibnamefont{Berger}},
  \bibinfo{journal}{Phys. Rev. B} \textbf{\bibinfo{volume}{89}},
  \bibinfo{pages}{060401} (\bibinfo{year}{2014}).

\bibitem[{\citenamefont{K\"obler et~al.}(1999)\citenamefont{K\"obler, Hoser,
  Kawakami, Chatterji, and Rebizant}}]{Kobler1999}
\bibinfo{author}{\bibfnamefont{U.}~\bibnamefont{K\"obler}},
  \bibinfo{author}{\bibfnamefont{A.}~\bibnamefont{Hoser}},
  \bibinfo{author}{\bibfnamefont{M.}~\bibnamefont{Kawakami}},
  \bibinfo{author}{\bibfnamefont{T.}~\bibnamefont{Chatterji}},
  \bibnamefont{and} \bibinfo{author}{\bibfnamefont{J.}~\bibnamefont{Rebizant}},
  \bibinfo{journal}{Journal of Magnetism and Magnetic Materials}
  \textbf{\bibinfo{volume}{205}}, \bibinfo{pages}{343 } (\bibinfo{year}{1999}).

\bibitem[{\citenamefont{Kobets et~al.}(2010)\citenamefont{Kobets, Dergachev,
  Khatsko, Rykova, Lemmens, Wulferding, and Berger}}]{Kobets2010}
\bibinfo{author}{\bibfnamefont{M.~I.} \bibnamefont{Kobets}},
  \bibinfo{author}{\bibfnamefont{K.~G.} \bibnamefont{Dergachev}},
  \bibinfo{author}{\bibfnamefont{E.~N.} \bibnamefont{Khatsko}},
  \bibinfo{author}{\bibfnamefont{A.~I.} \bibnamefont{Rykova}},
  \bibinfo{author}{\bibfnamefont{P.}~\bibnamefont{Lemmens}},
  \bibinfo{author}{\bibfnamefont{D.}~\bibnamefont{Wulferding}},
  \bibnamefont{and} \bibinfo{author}{\bibfnamefont{H.}~\bibnamefont{Berger}},
  \bibinfo{journal}{Low Temperature Physics} \textbf{\bibinfo{volume}{36}},
  \bibinfo{pages}{176} (\bibinfo{year}{2010}).

\bibitem[{\citenamefont{Laurita et~al.}(2015)\citenamefont{Laurita,
  Deisenhofer, Pan, Morris, Schmidt, Johnsson, Tsurkan, Loidl, and
  Armitage}}]{Laurita2015}
\bibinfo{author}{\bibfnamefont{N.~J.} \bibnamefont{Laurita}},
  \bibinfo{author}{\bibfnamefont{J.}~\bibnamefont{Deisenhofer}},
  \bibinfo{author}{\bibfnamefont{L.}~\bibnamefont{Pan}},
  \bibinfo{author}{\bibfnamefont{C.~M.} \bibnamefont{Morris}},
  \bibinfo{author}{\bibfnamefont{M.}~\bibnamefont{Schmidt}},
  \bibinfo{author}{\bibfnamefont{M.}~\bibnamefont{Johnsson}},
  \bibinfo{author}{\bibfnamefont{V.}~\bibnamefont{Tsurkan}},
  \bibinfo{author}{\bibfnamefont{A.}~\bibnamefont{Loidl}}, \bibnamefont{and}
  \bibinfo{author}{\bibfnamefont{N.~P.} \bibnamefont{Armitage}},
  \bibinfo{journal}{Phys. Rev. Lett.} \textbf{\bibinfo{volume}{114}},
  \bibinfo{pages}{207201} (\bibinfo{year}{2015}).

\bibitem[{\citenamefont{Sparks et~al.}(1961)\citenamefont{Sparks, Loudon, and
  Kittel}}]{Sparks1961}
\bibinfo{author}{\bibfnamefont{M.}~\bibnamefont{Sparks}},
  \bibinfo{author}{\bibfnamefont{R.}~\bibnamefont{Loudon}}, \bibnamefont{and}
  \bibinfo{author}{\bibfnamefont{C.}~\bibnamefont{Kittel}},
  \bibinfo{journal}{Phys. Rev.} \textbf{\bibinfo{volume}{122}},
  \bibinfo{pages}{791} (\bibinfo{year}{1961}).

\bibitem[{\citenamefont{Schl\"omann}(1961)}]{Schlomann1961}
\bibinfo{author}{\bibfnamefont{E.}~\bibnamefont{Schl\"omann}},
  \bibinfo{journal}{Phys. Rev.} \textbf{\bibinfo{volume}{121}},
  \bibinfo{pages}{1312} (\bibinfo{year}{1961}).

\bibitem[{\citenamefont{Kittel and Abrahams}(1953)}]{Kittel1953}
\bibinfo{author}{\bibfnamefont{C.}~\bibnamefont{Kittel}} \bibnamefont{and}
  \bibinfo{author}{\bibfnamefont{E.}~\bibnamefont{Abrahams}},
  \bibinfo{journal}{Rev. Mod. Phys.} \textbf{\bibinfo{volume}{25}},
  \bibinfo{pages}{233} (\bibinfo{year}{1953}).

\bibitem[{\citenamefont{Zhitomirsky and Chernyshev}(2013)}]{Zhitomirsky2013}
\bibinfo{author}{\bibfnamefont{M.~E.} \bibnamefont{Zhitomirsky}}
  \bibnamefont{and} \bibinfo{author}{\bibfnamefont{A.~L.}
  \bibnamefont{Chernyshev}}, \bibinfo{journal}{Rev. Mod. Phys.}
  \textbf{\bibinfo{volume}{85}}, \bibinfo{pages}{219} (\bibinfo{year}{2013}).

\bibitem[{\citenamefont{Chernyshev}(2012)}]{Chernyshev2012}
\bibinfo{author}{\bibfnamefont{A.~L.} \bibnamefont{Chernyshev}},
  \bibinfo{journal}{Phys. Rev. B} \textbf{\bibinfo{volume}{86}},
  \bibinfo{pages}{060401} (\bibinfo{year}{2012}).

\bibitem[{\citenamefont{Oguchi}(1960)}]{Oguchi1960}
\bibinfo{author}{\bibfnamefont{T.}~\bibnamefont{Oguchi}},
  \bibinfo{journal}{Phys. Rev.} \textbf{\bibinfo{volume}{117}},
  \bibinfo{pages}{117} (\bibinfo{year}{1960}).

\bibitem[{\citenamefont{Rezende}(2009)}]{Rezende2009}
\bibinfo{author}{\bibfnamefont{S.~M.} \bibnamefont{Rezende}},
  \bibinfo{journal}{Phys. Rev. B} \textbf{\bibinfo{volume}{79}},
  \bibinfo{pages}{174411} (\bibinfo{year}{2009}).

\bibitem[{\citenamefont{Ord\'o\~nez Romero
  et~al.}(2009)\citenamefont{Ord\'o\~nez Romero, Kalinikos, Krivosik, Tong,
  Kabos, and Patton}}]{Ordonez2009}
\bibinfo{author}{\bibfnamefont{C.~L.} \bibnamefont{Ord\'o\~nez Romero}},
  \bibinfo{author}{\bibfnamefont{B.~A.} \bibnamefont{Kalinikos}},
  \bibinfo{author}{\bibfnamefont{P.}~\bibnamefont{Krivosik}},
  \bibinfo{author}{\bibfnamefont{W.}~\bibnamefont{Tong}},
  \bibinfo{author}{\bibfnamefont{P.}~\bibnamefont{Kabos}}, \bibnamefont{and}
  \bibinfo{author}{\bibfnamefont{C.~E.} \bibnamefont{Patton}},
  \bibinfo{journal}{Phys. Rev. B} \textbf{\bibinfo{volume}{79}},
  \bibinfo{pages}{144428} (\bibinfo{year}{2009}).

\bibitem[{\citenamefont{Kreisel et~al.}(2009)\citenamefont{Kreisel, Sauli,
  Bartosch, and Kopietz}}]{Kreisel2009}
\bibinfo{author}{\bibfnamefont{A.}~\bibnamefont{Kreisel}},
  \bibinfo{author}{\bibfnamefont{F.}~\bibnamefont{Sauli}},
  \bibinfo{author}{\bibfnamefont{L.}~\bibnamefont{Bartosch}}, \bibnamefont{and}
  \bibinfo{author}{\bibfnamefont{P.}~\bibnamefont{Kopietz}},
  \bibinfo{journal}{The European Physical Journal B}
  \textbf{\bibinfo{volume}{71}}, \bibinfo{pages}{59} (\bibinfo{year}{2009}).

\bibitem[{\citenamefont{Wigen}(1994)}]{Wigen1994}
\bibinfo{author}{\bibfnamefont{P.~E.} \bibnamefont{Wigen}},
  \emph{\bibinfo{title}{Nonlinear Phenomena and Chaos In Magnetic Materials}}
  (\bibinfo{publisher}{World Scientific}, \bibinfo{year}{1994}).

\end{thebibliography}

\end{document}